\RequirePackage{fix-cm}
\documentclass{svjour3}                     
\smartqed  
\usepackage[]{natbib}
\usepackage[utf8x]{inputenc}
\usepackage{graphicx}
\usepackage{amsmath,amssymb}
\usepackage{multirow}
\usepackage{marvosym}
\usepackage{latexsym}
 \journalname{}
\begin{document}

\title{A novel computational modelling to describe the anisotropic, remodelling and reorientation behaviour of collagen fibres in articular cartilage
}
\subtitle{}

\titlerunning{A novel computational modelling to describe the anisotropic, remodelling and reorientation behaviour}        

\author{S. Cortez        \and
        A. Completo  \and
				J. L. Alves
}


\institute{S. Cortez ($\Letter$) \and J. L. Alves \at
              Department of Mechanical Engineering\\
							University of Minho\\
              Guimarães, Portugal\\
              \email{{scortez,jlalves}@dem.uminho.pt}           
           \and
           A. Completo \at
              Department of Mechanical Engineering\\
							University of Aveiro\\
							Aveiro, Portugal
}

\date{Received: date / Accepted: date}

\maketitle
\begin{abstract}
In articular cartilage the orientation of collagen fibres is not uniform varying mostly with the depth of the tissue. Besides, the biomechanical response of each layer of the articular cartilage differs from the neighbouring ones, evolving through thickness as a function of the distribution, density and orientation of the collagen fibres. Based on a finite element implementation, a new continuum formulation is proposed to describe the remodelling and reorientation behaviour of the collagen fibres under arbitrary mechanical loads. The cartilaginous tissue is modelled based on a hyperelastic formulation, being the ground isotropic matrix described by a neo-Hookean law and the fibrillar anisotropic part modelled by a new anisotropic formulation introduced for the first time in the present work, in which both reorientation and remodelling are taken into account. To characterize the orientation of fibres, a structure tensor is defined to represent the expected distribution and orientation of fibres around a reference direction. The isotropic and anisotropic constitutive parameters were determined by the good validation of the numerical models with the experimental data available from the literature. Considering the effect of realistic collagen fibre reorientation in the cartilage tissue, the remodelling algorithm associated with a distribution of fibres model showed accurate results with few numerical calculations.

\end{abstract}

\section{Introduction \label{intro}}
In tissue engineering, the mechanical properties of the engineered material are crucial for any application and/or treatment. The mechanical stimuli imposed on bioreactors are between the most important factors on the tissue growth. Finite element analysis based on computer models for articular cartilage is commonly used to understand the intrinsic growth factors (collagen fibres orientation, solute transport, growth of extracellular matrix constituents and others) and to simulate the effect of changes of the tissue structure under well-defined loading conditions \citep{chung2010analysis,nava2013multiphysics, hossain2014prediction,bandeiras2015cartilage}. In order to improve the mechanical properties of the tissue-engineered cartilage, the organization of the collagen network associated with the mechanical loading needs a deeper attention. \\
The collagen fibres in the articular cartilage have a non-uniform orientation and distribution varying with the depth of the tissue. Because of this inhomogeneous structure, each cartilage layer responds differently to the same mechanical load \citep{pierce2013hyperelastic, federico2008towards, khoshgoftar2013effect, gasser2006hyperelastic, federico2010nonlinear}. The role of fibres is essentially mechanical, promoting tissue stiffness and strength. Besides, the mechanical behaviour of connective tissues is strongly influenced by the structural arrangement of the fibres. Consistent with classical arcade-like descriptions of collagen network organization in cartilage \citep{wilson2004stresses}, fibres in the superficial zone are preferably oriented in the plane parallel to the articular surface, while in the middle zone fibres are dispersed with a random orientation.  In the deep tissue zone, fibres tend to be oriented perpendicularly to the bone cartilage interface, i.e., perpendicular to the articular surface \citep{pearle2005basic, wilson2007depth}. \\
Several nonlinear constitutive models have been proposed to study the mechanical behaviour of tissues containing collagen fibres and some of constitutive laws consider a negligible resistance ('buckling') of fibres under compression \citep{holzapfel2000new,federico2008towards}. 
Finite element (FE) models have been developed for families of fibres oriented in different directions \citep{holzapfel2000new, holzapfel2001viscoelastic}. The constitutive approach based on angular integrals requires extensive calculations \citep{ateshian2009modeling}. Based on a hyperelastic free-energy function, these models have been improved using a generalized structure tensor that characterizes the statistical dispersed collagen fibre orientation \citep{gasser2006hyperelastic}.  Still, even if this approach requires a small number of calculations to obtain the strain and stresses fields, these models are limited \citep{cortes2010characterizing}. \citet{federico2008towards} proposed a constitutive model represented by an integral form of the elastic strain energy potential, which is performed on the unit sphere. Here, the integrals evaluation is performed using the spherical designs methods \citep{pierce2015microstructurally}. Recent progress in FE models has improved our understanding of depth-dependent properties of cartilage incorporating the fibre response using a reference direction and a fibre dispersion around this direction \citep{pierce2013hyperelastic,wilson2007depth}. Theoretical models to predict the collagen network architecture in various soft tissues assumes that collagen fibres align along the preferred fibre directions that are situated between the positive principal strain directions \citep{driessen2003remodelling, wilson2006prediction, driessen2008remodelling}. The collagen alignment has been shown to align with respect to loading in articular cartilage, predicting the development of the typical Benninghoff-type collagen fibre orientation \citep{wilson2004stresses,khoshgoftar2011mechanical}.
\subsection{Aim of the study \label{aim}}
In the present study, a novel continuum anisotropic hyperelastic formulation taking into account both fibres reorientation and fibres remodelling is proposed, aiming to investigate the role and evolution of the depth dependent collagen network in the articular cartilage. A structural tensor associated with the dispersion of the embedded collagen fibres is introduced. The arrangement of the fibres is not represented by a numerical integration on the sphere surface (with a spherical harmonic distribution), but with a distribution of fibres simply defined around a reference direction and an ellipsoidal distribution assumed. The aim is to evaluate whether the new remodelling algorithm based on \citet{wilson2006prediction}, associated with this fibre-reinforced FE model is consistent with the orientation and distribution of the collagen fibres observed in native articular cartilage. The remodelling algorithm is therefore envisaged to be a valuable tool for the development of improved loading protocols for the tissue engineering of articular cartilage, and the understanding of structural adaptation of the collagen network during random loading conditions.
\section{Continuum mechanical framework \label{framework}}
\subsection{Basic Kinematics \label{kinematics}}
Let ${\Omega _0} \subset {\mathbb{R}^3}$ be the (fixed) reference configuration of a continuous body. The body undergoes a deformation ${\bf{\chi }}$, which transforms a typical reference material point ${\bf{X}} \in {{\rm{\Omega }}_{\rm{0}}}$ into a spatial point ${\bf{x}} = \chi \left( {\bf{X}} \right) \in {\rm{\Omega }}$ in the deformed configuration. Let ${\bf{F}}({\bf{X}}) = \partial \chi \left( {\bf{X}}
\right)/\partial {\bf{X}}$ be the deformation gradient and ${J}\left(\bf{X}\right) = \det{\bf{F}}({\bf{X}})$ the local volumetric deformation ratio. Due to the material incompressibility of soft biological tissues (such as articular cartilage), the deformation gradient can be decomposed into a spherical (dilatational) and a unimodular (distortional) part as:
\begin{equation}
{\bf{F}} = \left( {{J^{1/3}}{\bf{I}}} \right){\bf{\bar F}}
\label{eq01}	
\end{equation}
where ${\bf{I}}$ is the second order unity tensor and ${\bf{\bar F}}$ is associated to the part of the total deformation gradient that does not produce any change of volume.
\begin{figure*}[ht]
	\centering
		\includegraphics[width=1.0\textwidth]{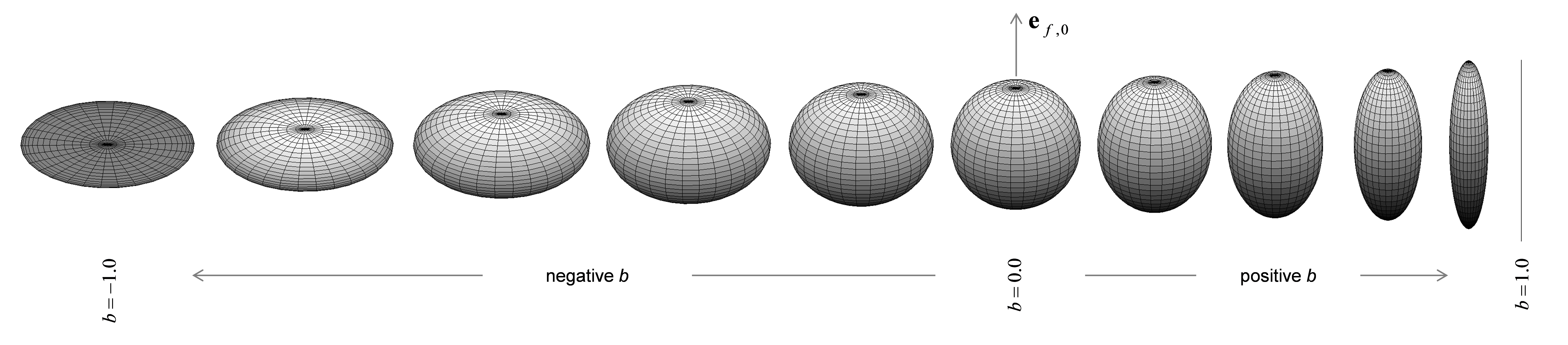}
	\caption{Three-dimensional graphical representation of the ellipsoidal distribution of fibres given a reference direction ${\vec e_{f,0}}$.}
	\label{fig:fig1}
\end{figure*}   
Soft tissues (such as cartilage) have a non-linear elastic mechanical behaviour. Generally, soft tissues like cartilage are modelled with pseudo-elastic or hyperelastic models. Adopting a hyperelastic formulation, the Lagrangian second Piola-Kirchhoff stress tensor (${\bf{\Pi }}$) can be derived from the strain energy density potential as
\begin{equation}
{\bf{\Pi }} = \frac{{\partial W}}{{\partial {\bf{E}}}} = 2\frac{{\partial W\left( {\bf{C}} \right)}}{{\partial {\bf{C}}}}
\label{eq02}	
\end{equation}
where ${\bf{C}} = {{\bf{F}}^{\bf{T}}}{\bf{F}} = {J^{2/3}}{\bf{\bar C}}$ and ${\bf{E}} = \frac{1}{2}\left( {{\bf{C}} - {\bf{I}}} \right)$ are the right Cauchy-Green strain tensor and the Green-Lagrange strain tensor, respectively. The scalar strain energy function ($W$) depends on a set of invariants of ${\bf{C}}$. This potential can also be decomposed into volumetric (${W_v}$) and isochoric ($\bar W$) parts,
\begin{equation}
W\left( {J,{\bf\bar{C}}} \right) = {W_v}\left( J \right) + \bar W\left( {\bf\bar{C}},... \right)
\label{eq03}	
\end{equation}
being the first term associated with the volumetric change and the second one with the isochoric deformation \citep{alves2010numerical,castro2014disc}.

In case of an inhomogeneous structure like articular cartilage, the mechanical response is driven by either the matrix or, mainly, by the collagen fibres. While the matrix determines the isotropic response, the collagen fibres determine the anisotropic response of the articular tissue.  Therefore, the total elastic strain energy can be decomposed into the elastic energy associated to the deformation of the matrix (i.e. the non-fibrous isotropic part) plus the elastic energy associated to the deformation of the collagen fibres (i.e. the fibrous anisotropic part), such that:
\begin{equation}
\bar W\left( {{\bf\bar{C}},{\bf{H}}} \right) = {\bar W_{iso}}\left( {{\bf{\bar C}}} \right) + {\bar W_{aniso}}\left( {{\bf\bar{C}},{\bf{H}}} \right)
\label{eq04}	
\end{equation}

In the present work, the isotropic material is described by the neo-Hookean law, 
\begin{equation}
{\bar W_{iso}}({\bf{\bar C}}) = \frac{\mu }{2}\left( {{{\bar I}_1} - 3} \right)
\label{eq05}	
\end{equation}
and the anisotropic fibrous material by a new fibre-reinforced model proposed in this work, based on \citet{gasser2006hyperelastic} and \citet{holzapfel2010constitutive}, and on a new structure tensor ${\bf{H}}$,
\begin{equation}
{\bar W_{aniso}}({\bf{C}},{\bf{H}}) = \frac{{{k_1}}}{{2{k_2}}}\left[ {\exp ({k_2}{\langle{E_1}\rangle}^2) - 1} \right]
\label{eq06}	
\end{equation}
where $\mu$ is the shear modulus of the isotropic matrix of the articular cartilage, ${\bar I_1}$ is the first invariant of ${\bf{C}}$, ${\bf{H}}$ is the general structure tensor, and $k_1$ and $k_2$ are material parameters associated with fibres, which can be determined by experimental validation. ${E_1}$ is a new pseudo-invariant defined as
\begin{equation}
{E_1} = {\bf{H}}:{\bf{C}} - {\bf{H}}:{\bf{I}}
\label{eq07}	
\end{equation}

The new general structure tensor ${\bf{H}}$ plays a major role in the definition of this new pseudo-invariant ${E_1}$. Indeed, the role of the structure tensor is to take into account any physically based distribution of the collagen fibres around a reference direction ${\vec{e}_{f,0}}$, as can be seen in the different layers of the articular cartilage. Therefore, the structure tensor ${\bf{H}}$ depends on ${b}$  parameter, which defines the dispersion of fibres around a random reference direction ${\vec{e}_{f,0}}$. The compact form for structure tensor can be given by:
\begin{equation}
{{\bf{H}}^ -} = \left({1+{\beta^-}{{\left| b \right|}^\alpha }} \right)\;{\bf{I}}-{\left| b \right|^\alpha }\left( {1 + {\beta ^-}} \right)\;{\vec e_{f,0}} \otimes \;{\vec e_{f,0}}\\
\label{eq08}
\end{equation}
with $b \in \left[ {-1,0} \right[$,

\begin{equation}
{{\bf{H}}^ + } = \left( {1-{b^\alpha }}\right)\;{\bf{I}}+{b^\alpha}\left({1+{\beta ^ + }} \right)\;{\vec e_{f,0}} \otimes \;{\vec e_{f,0}} \\
\label{eq09}
\end{equation}
with $b \in \left[ {0,1} \right]$.
\\
\\
The parameters $\alpha$ and $\beta$ are material constants, varying between $\left[-\infty,+\infty\right]$, aimed at increasing the flexibility of the model in order to allow a best fit of the nonlinear structure tensor stress response depending on parameter $b$ to experimental data. In this new formulation, the range of $b$ parameter is defined between $-1$ and $1$. The following well-defined particular cases can be identified: $a)$ lower limit $b=-1$ describes an isotropic distribution in the plane normal of the reference direction ${\vec e_{f,0}}$ (i.e., the superficial zone of the articular cartilage); $b)$ the parameter $b=0$ corresponds to the three-dimensional isotropic arrangement of the fibrous network (i.e., the middle layers of the articular cartilage); and $c)$ the upper limit defines the alignment of all collagen fibres with the reference direction ${\vec e_{f,0}}$ (i.e., the deep zone). Fig. \ref{fig:fig1} shows a graphical representation of the fibres distribution with a particular reference direction ${{\vec e}_{f,0}}$.
Table \ref{tab:1} shows the values of the distribution parameter $b$, a schematic representation of the orientation of fibres and the structure tensor for each typical zone of the articular cartilage assuming the reference direction given by ${\vec e}_{f,0}\equiv z$. 

A final remark concerning the usage of the Macaulay brackets (defined by the operator $\left\langle \cdot\right\rangle$) in Eq. \ref{eq06}. A common assumption is that fibres can only be loaded in tension, and suffer buckling in compression \citep{federico2008towards}. In order to take this into account, the elastic energy associated to the compression of the fibres must be neglected, and thus if  ${E_1}\leq0$, the term inside the Macaulay brackets becomes zero and the energy function is reduced to the isotropic part. 
\subsection{Collagen remodelling \label{remodelling}}
The remodelling of the collagen network in the articular cartilage in general, and particularly in tissue engineered, occurs at two levels: on the one hand, the evolution of the collagen density, which determines the strength of the tissue; on the other hand, the reorientation of the collagen fibres network, which determines the anisotropy of the tissue. Together, density and orientation, as well as remodelling and reorientation, determine the mechanical response of the tissue to arbitrary mechanical loads. 
In what follows is presented the model proposed to describe both phenomena, i.e. the collagen fibres remodelling and reorientation. Knowing that collagen fibres growth and remodelling is due to the mechanical stimuli, and thus due to tissue deformation and strain fields, let us consider that ${\vec e_1}>{\vec e_2}>{\vec e_3}$, the principal vectors (eigenvectors), or principal strain directions, of the Green-Lagrange  strain tensor ${\bf{E}}$, and ${\lambda_1}>{\lambda_2}>{\lambda_3}$ the principal values (eigenvalues) of the same strain tensor. The principal values can be either positive or negative, representing so a tensile or a compressive strain state, respectively, with respect to the correspondent principal strain vector \citep{driessen2003remodelling,wilson2006prediction,khoshgoftar2011mechanical}.

\begin{figure}[h]
	\centering
		\includegraphics[width=0.3\textwidth]{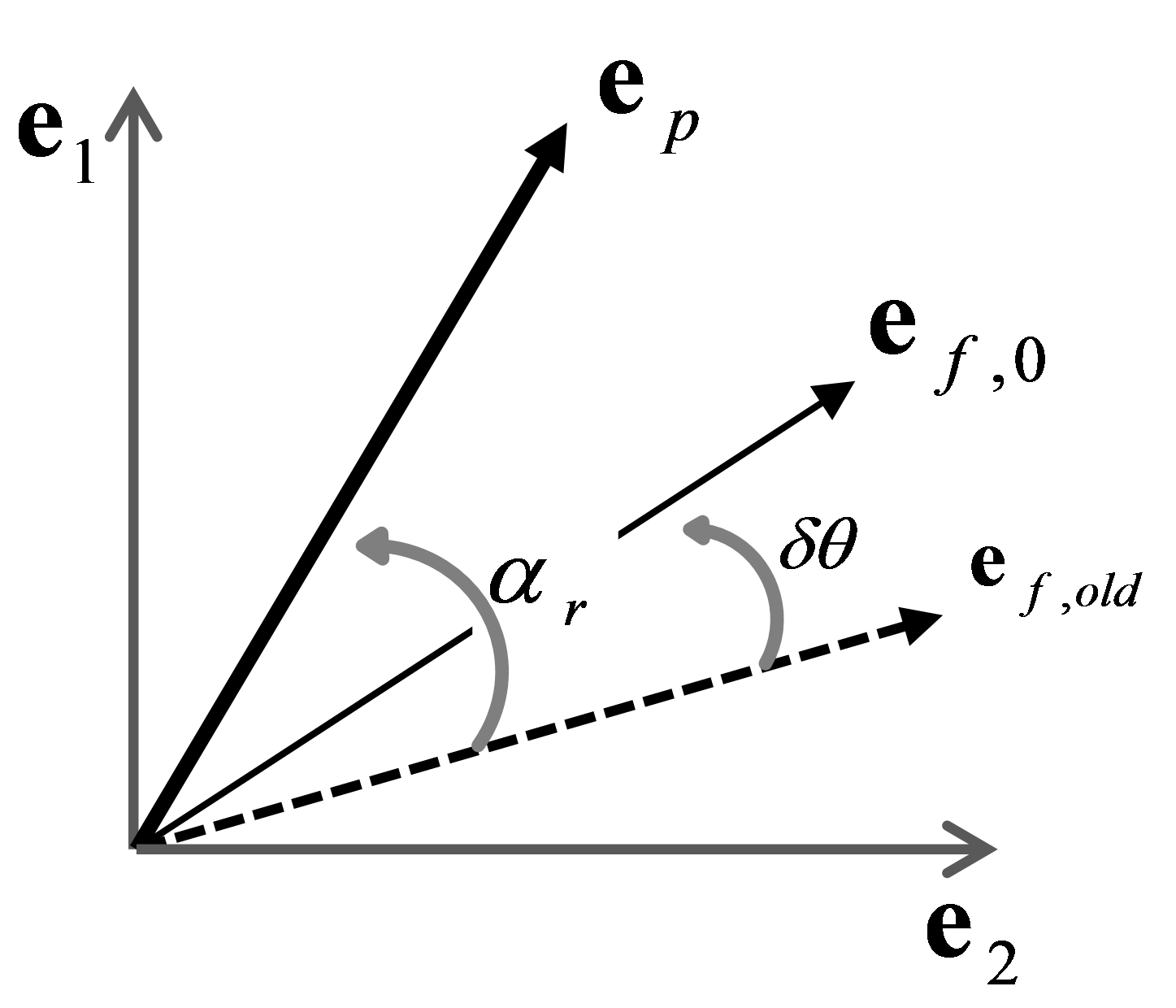}
	\caption{Schematic representation of the preferred fibre direction ${{\vec e}_{p}}$ situated in between the positive principal strain directions ${{\vec e}_{1}}$ and ${{\vec e}_{2}}$. Note that vector ${{\vec e}_{3}}$ is perpendicular to the plane. The fibre direction with respect to the undeformed configuration ${{\vec e}_{f,0,old}}$ is rotated toward the preferred fibre direction ${{\vec e}_{p}}$ over an angle d${\theta}$ resulting in the new fibre direction ${{\vec e}_{f,0}}$. $\alpha_r$ denotes the angle between ${{\vec e}_{p}}$ and ${{\vec e}_{f,0}}$.}
	\label{fig:fig2}
\end{figure}

Based on the hypothesis that collagen fibres align and growth along the positive principal strain directions, i.e. along the tensile mechanical stimulus,
 the fibre alignment is done through preferential direction ${\vec e_p}$ (see Fig. \ref{fig:fig2}) defined by:
\begin{equation}
{\vec e_p} = {\vec e_3}
\label{eq10}	
\end{equation} 
when the principal value associated to a given principal vector is in compression and two principal values associated to the other two principal vectors are in tension (i.e. ${\vec e_1},{\vec e_2}>0$ and ${\vec e_3}<0$) , or defined by
\begin{equation}
{\vec e_p} = {\vec e_1}
\label{eq11}	
\end{equation} 
when two axes are in compression and an axis in tension (${\vec e_1},{\vec e_2}<0$ and ${\vec e_3}>0$). ${\vec e_1},{\vec e_2}$ and ${\vec e_3}$ are the eigenvectors of the right-Cauchy tensor ${\bf C}$, i.e., they are the principal directions of the strain. 
The collagen fibres tend to reorient toward the preferred fibre direction with a given angular velocity. Thus, the reorientation model defines the rate at which the reference axis ${{\vec e}_{f,0}}$ of the tissue rotates to the preferential direction ${{\vec e}_{p}}$:
\begin{equation}
\frac{{d\theta }}{{dt}} = \kappa {\alpha _r} = \kappa \arccos \left\| {{{\vec e}_{f,0}} \cdot {{\vec e}_p}} \right\|
\label{eq12}	
\end{equation} 
where ${\alpha _r}$ is the angle between the current reference fibre direction ${{\vec e}_{f,0}}$ (undeformed configuration) and the preferred fibre direction ${{\vec e}_p}$ determined at a given instant as a function of the local strain field (see Fig. \ref{fig:fig2}). 

To control the rate of reorientation, a positive constant $\kappa$ was defined. The fibres are rotated around the following rotation axis,
\begin{equation}
{\vec e_r} = \frac{{{{\vec e}_{f,0}} \otimes {{\vec e}_p}}}{{\left\| {{{\vec e}_{f,0}} \otimes {{\vec e}_p}} \right\|}}
\label{eq13}	
\end{equation} 
However, future experimental results shall contribute to improve our understanding about fibre reorientation.

The collagen fibres will not only reorient but also redistribute around the reference direction. The distribution of the collagen fibres around the reference direction is assumed to be described by parameter $b$ and modelled by the structure tensor $\bf{H}$. Adopting for parameter $b$ a similar evolution law as adopted for the evolution of the reference direction, the following equation can be written as follow, 
\begin{equation}
\frac{db}{dt}=r_b\left (b_t-b_0 \right)
\label{eq14}	
\end{equation} 
where $b_0$ is the current value of the distribution parameter $b$ and $b_t$ is the target value of distribution of fibres, to be calculated from the strain field and strain loading history. The reorientation rate of distribution is defined by $r_b$ and it is associated with the regeneration process of the cartilage. To establish the natural phenomenological process of the tissue, we hypothesized that ($\textit{i}$) in compression reaches the value of $-1$, i.e., the fibres tend to be oriented in the direction perpendicular to the fibre reference direction, and ($\textit{ii}$) in tension fibres tend to the positive value of $+1$, where the fibres are perfectly aligned with the fibre reference direction ${{\vec e}_{f,0}}$ (see Fig. \ref{fig:fig2}).
\begin{table*}[ht]
\caption{Distribution parameter, schematic representation of the fibre orientation and structure tensor for superficial, middle and deep zone of cartilage for a fibres reference direction of ${{\vec e}_{f,0}}$.}
\label{tab:1}       
\centering
\renewcommand{\arraystretch}{1.5}
\begin{tabular}{ l c c c }
\hline
Zone                                             & Superficial & Middle & Deep \\ \hline
$b$                                              & -1.0        & 0.0    & 1.0  \\
\multirow{6}{*}{Fibre orientation} & & & \\
 & \includegraphics[width=0.2\textwidth]{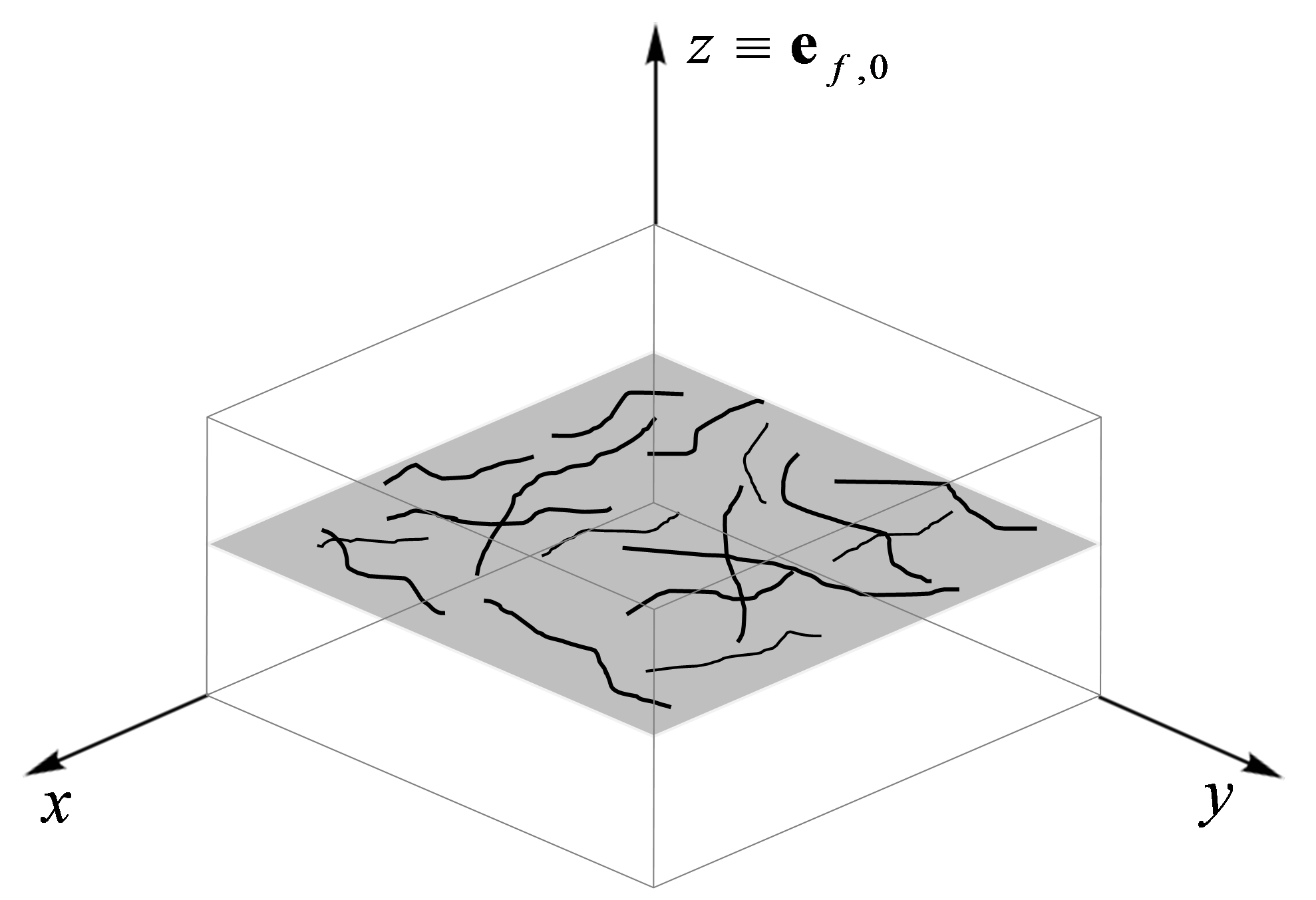} & \includegraphics[width=0.2\textwidth]{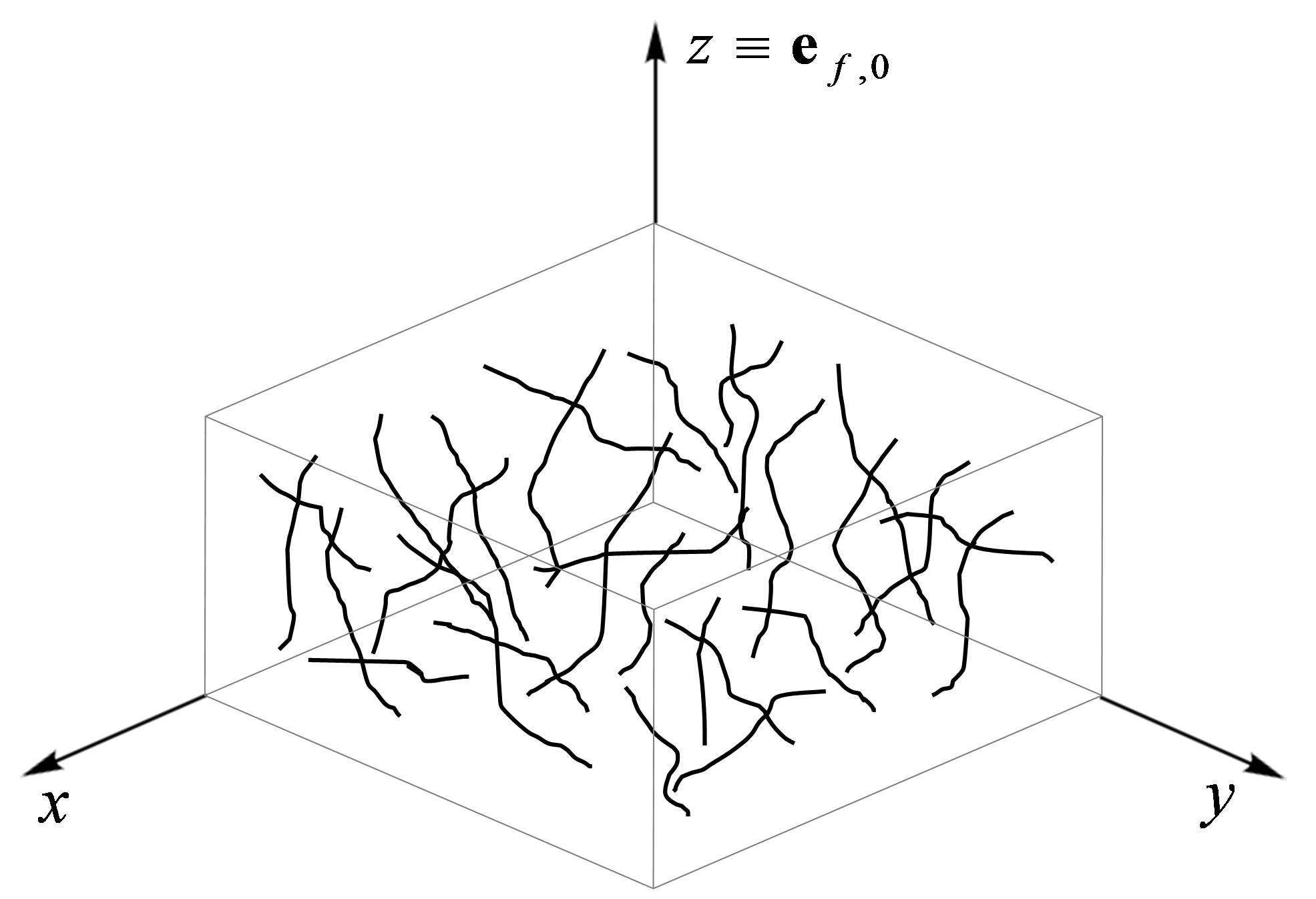} & \includegraphics[width=0.2\textwidth]{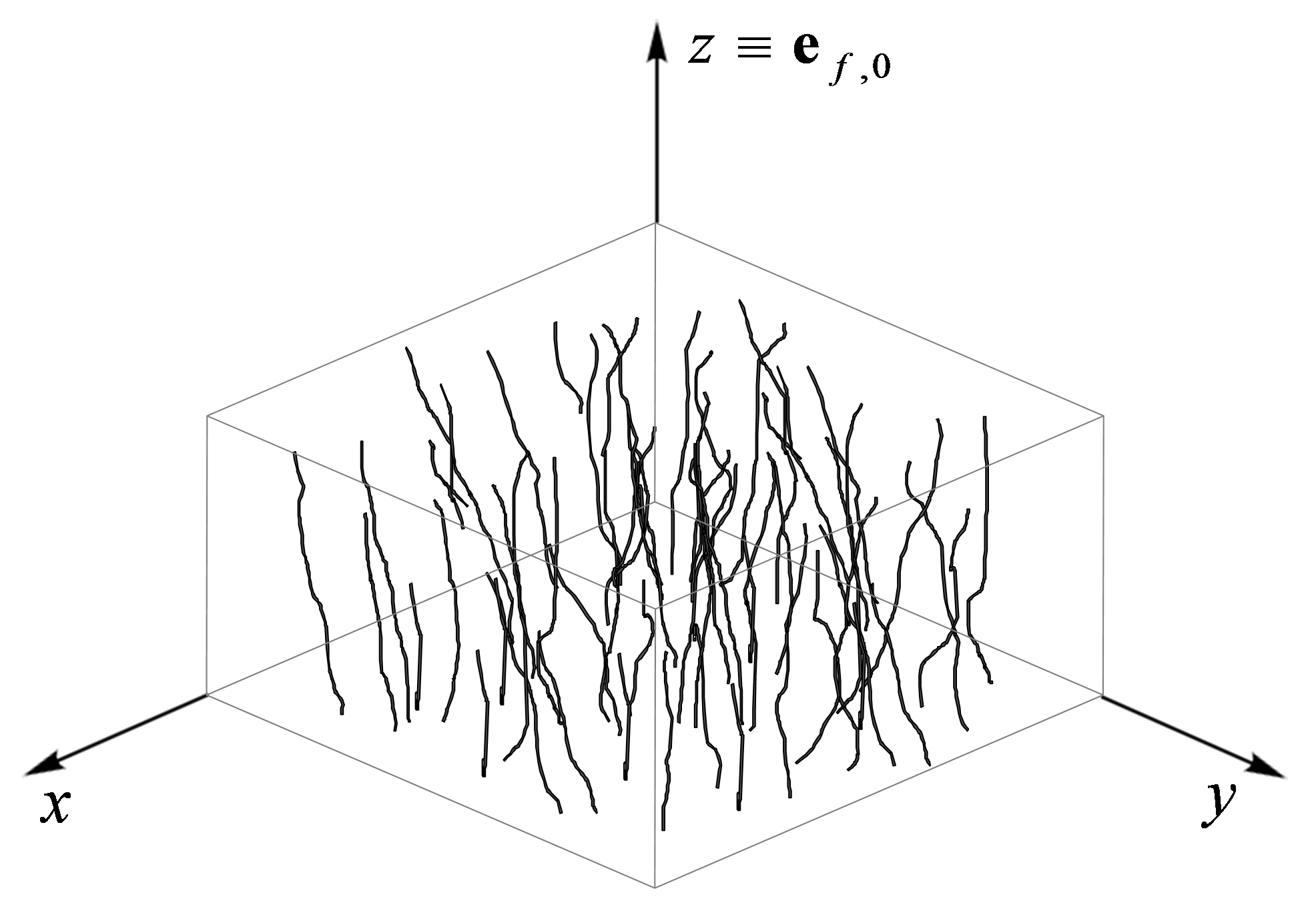}  \\
 & & & \\
\multirow{2}{*}{\begin{tabular}[c]{@{}l@{}}Structure tensor in tension\end{tabular}}     & $\left[ \begin{matrix} 0 &  & \\  & 0 & \\  &  & 1-\left|b^{\alpha}\right|\end{matrix}\right]$            & $\left[ \begin{matrix} 0 & & &&&& \\ &&&& 0 & &&&& \\ &&&& & & & & 1  \end{matrix}\right]$   & $\left[ \begin{matrix} 0 &  & \\  & 0 & \\  &  & 1+\beta^{+}b^{\alpha}\end{matrix}\right]$       \\
                                                    &  \multicolumn{3}{c}{} \\
																										& & & \\
\multirow{2}{*}{\begin{tabular}[c]{@{}l@{}}Structure tensor in compression\end{tabular}}  &  $\left[ \begin{matrix} 1+\beta^{-}\left|b^{\alpha}\right| &  & \\  & 1+\beta^{-}\left|b^{\alpha}\right| & \\  &  & 0 \end{matrix}\right]$            & $\left[ \begin{matrix}  1 & & &&&& \\ &&&& 1 & &&&& \\ &&&& & & & & 0 \end{matrix}\right]$       &  $\left[ \begin{matrix} 0 & & &&&& \\ &&&& 0 & &&&& \\ &&&& & & & & 0  \end{matrix}\right]$     \\
                                                    &  \multicolumn{3}{c}{}  \\  \hline
\end{tabular}
\end{table*}

\section{Finite element implementation \label{implementation}}
The validation of the proposed formulation and FE implementation includes three main steps. Firstly, an analytical procedure to adjust the material and structure parameters was employed, determining a set of material parameters for each cartilage zone. Using some literature studies \citep{ateshian1997finite,elliott2002direct,jurvelin2003mechanical,pierce2013hyperelastic}, the influence of these structure parameters were then investigated based on the nonlinear behaviour of the cartilage biphasic fibre-reinforced FE model. To explore the proposed remodelling algorithm, some numerical examples were implemented using the validated fibre-reinforced FE model, evaluating the performance of the collagen fibres reorientation that occurs in articular cartilage. At the end, a combination of all tissue layers in the same FE model was analysed. A total displacement corresponding to 10$\%$ of the specimen thickness was applied in the sample over a pseudo-time and different values of ${r_b}$ of 0.1, 0.2 and 0.3 (see Eq. \ref{eq13}) were evaluated.
\subsection{Loading and boundary conditions \label{boundary}}
Some studies and FE analysis in literature have been using a confined compression configuration to study the mechanical response of the collagen fibres in cartilage \citep{guo2015biphasic}. However, this boundary value problem exhibits some main drawbacks in terms of the response of the fibres under tensile loads, i.e., the fibres strain in the transverse direction to the main loading axis (compressive loading) will be null due to the confinement. In the superficial zone, this issue is more evident and constraining. Thus, to demonstrate that the proposed formulation and FE model can reproduce the experimental observations, unconfined compression and uniaxial tension problems were considered. 
\begin{figure}[h]
	\centering,
		\includegraphics[width=0.3\textwidth]{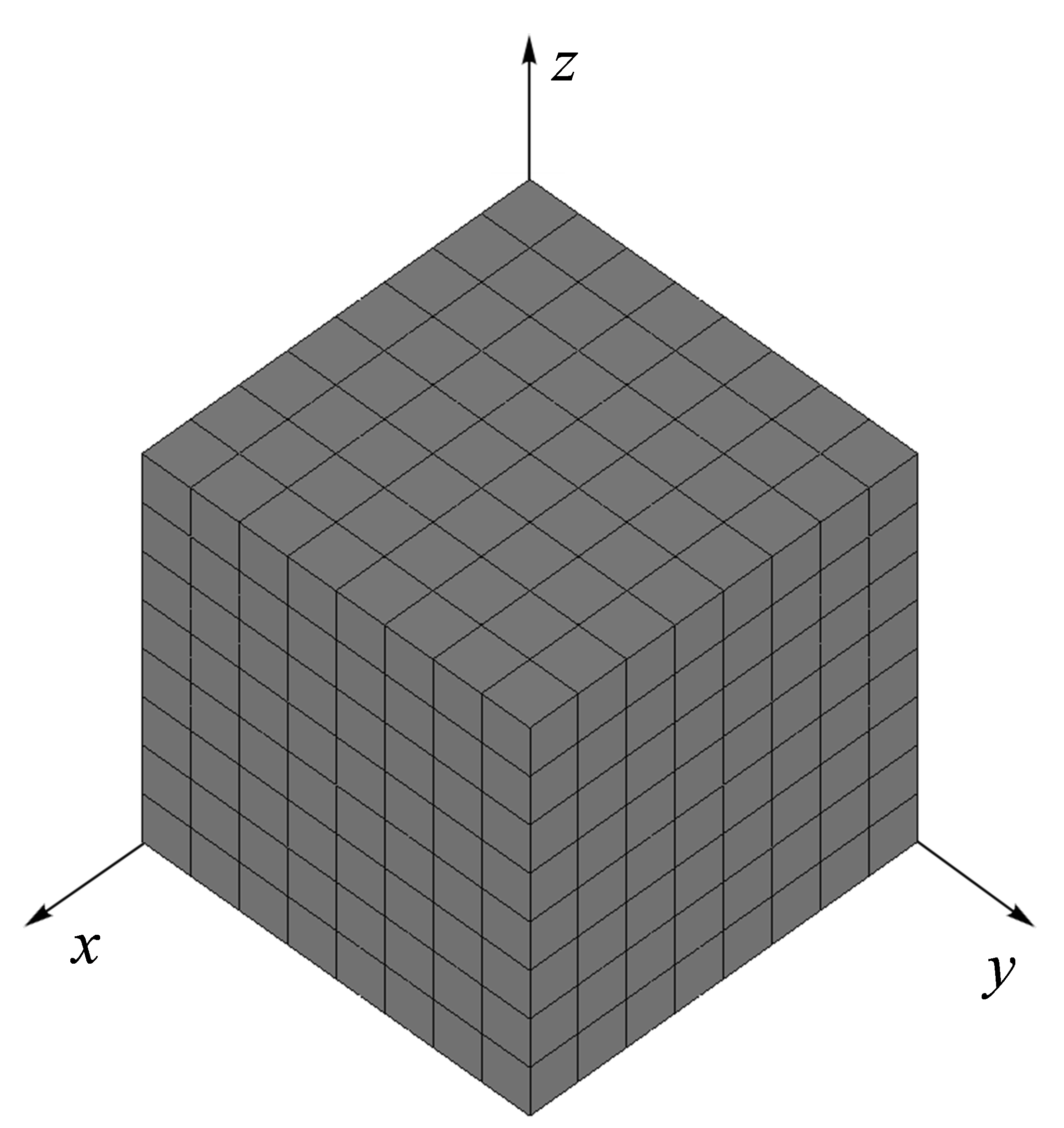}
	\caption{Cubic finite element mesh of a cartilage layer ($1.0$x$1.0$x$1.0$ mm$^3$) symmetrically constrained in $x0y$, $x0z$ and $y0z$ planes.}
	\label{fig:fig3}
\end{figure}
Both compression and tension simulations were performed using a cubic FE mesh ($1.0$x$1.0$x$1.0$ mm$^3$) discretized with $512$ quadratic 27-node hexahedral finite elements (see Fig. \ref{fig:fig3}). The FE model was symmetrically constrained in $x0y$, $x0z$ and $y0z$ planes and modelled on a home-developed open source FE solver. A fully-implicit Newton-Raphson iterative method to solve the nonlinear problem, and a biphasic formulation with total incompressibility of the solid matrix are used \citep{alves2010numerical,castro2014disc}.
An FE model of a cartilage cubic sample was built in unconfined compression for both superficial zone (with the fibres aligned parallel to articular surface) and middle zone (with the fibres randomly orientated). A displacement was applied to the upper surface of the sample up to an engineering strain of $-0.50$ and $-0.80$ for the superficial zone and middle zone, respectively.  As in the deep zone all fibres are in the same direction and perfectly aligned with the loading axis, the compressive response of the anisotropic part will be null due to the buckling assumption of the fibres under compression \citep{cortes2010characterizing}. Thus, the results in compression of this zone were not considered for analysis.
The uniaxial tension model was tested to validate the constitutive model with experimental curves for all cartilage zones (superficial, middle and deep). A displacement was applied to the upper surface of the sample up to an applied engineering strain of $+0.15$. In the superficial case, the loading axis is parallel to the plane in which all fibres are isotropically arranged, as represented in Fig. 4b, in opposition to the more biological orientation of the fibres with respect to the loading axis (see Fig. \ref{fig:fig4}). In this tensile condition, fibres are stretched along the loading direction and compressed along the perpendicular directions. Here, a different fibre reference direction  ${{\vec e}_{f,0}}$ was defined.

\begin{figure}[ht]
	\centering
		\includegraphics[width=0.5\textwidth]{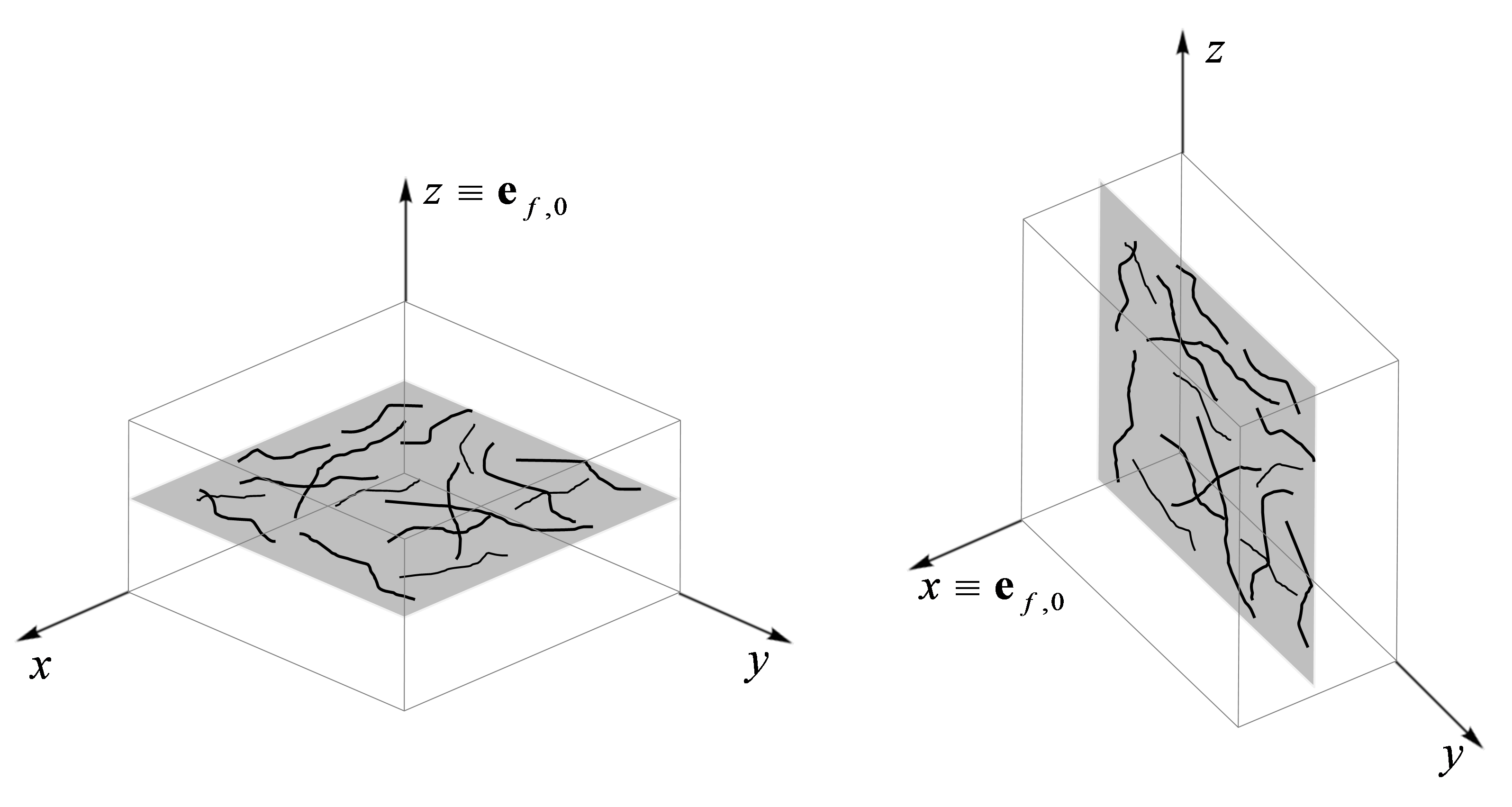}
	\caption{Orientation of the fibres in the superficial zone. Left:
scheme of the fibres distribution in the plane perpendicular to the reference direction ${\vec e_{f,0}}$. The loading direction coincides with the reference direction.  Right: scheme of the plane of fibre distribution aligned with the loading direction. Here, the reference direction of fibres was changed.}
	\label{fig:fig4}
\end{figure} 
\subsection{Material and structure data \label{parameters}}
Analysed some studies from literature, the isotropic material properties of the extracellular matrix were obtained. The shear modulus has been reported as constant for all zones in the articular cartilage. Despite the collagen content may influence the shear modulus \citep{pierce2013hyperelastic, responte2007collagens}, in this study this collagen content was not considered in the definition of the material properties. \\

The cartilage tension-compression nonlinear behaviour was analysed comparing the experimental data with the numerical results in order to find the structure parameters, i.e. parameters $\mu$, $k_1$, $k_2$, $alpha$, $beta+$ and $beta-$. For the sake of simplicity, the same numerical model was used in compression and in tension to simulate the superficial and middle zones. However, for the deep zone, only the uniaxial tension model was achieved, given that fibres will not respond under compression when they are aligned with the loading direction. Unfortunately, there are no experimental data to compare the response of the deep zone of the articular cartilage. Here, the same material parameters used in the simulation of $b=0$ were used for the simulation with $b=1$, representing the deep zone when the fibres are aligned with the reference fibre direction. For all examples, ${{\vec e}_{f,0}}$ was defined to be the same as the loading direction, except for the uniaxial tension of the superficial zone, where it was assumed as being perpendicular to the loading direction (see Fig. \ref{fig:fig4}). In this case, all fibres are dispersed in the perpendicular plane to the articular surface and the fibres are stretched in the loading direction. The fibre distribution parameter was defined as $b=-1$ for superficial zone (where the fibres are in the plane which is perpendicular to ${{\vec e}_{f,0}}$ and parallel to the articular surface), as $b=0$ for the middle zone (with a randomly fibre distribution) and $b=1$ for the deep zone (where all fibres are aligned with ${{\vec e}_{f,0}}$, i.e., perpendicular to the articular surface) \citep{wilson2006prediction, responte2007collagens,pierce2013hyperelastic}. 
\begin{table*}[ht]
\centering
\caption{Reference material, structure and remodelling parameters of the FE model characterizing cartilage layers in different loading conditions (C - compression and T - tension).}
\label{tab:2}       
\renewcommand{\arraystretch}{1.5}
\begin{tabular}{lclllllllllll}
\hline\noalign{\smallskip}
Zone                            & \begin{tabular}[c]{@{}l@{}}Loading \end{tabular}       &  ${{\vec e}_{f,0}}$         & $b$   & $\mu$(MPa)    & $k_1$(MPa)   & $k_2$   & $\alpha$ & $\beta+$ & $\beta-$ & $b_t$   & $\kappa$ & $r_b$\\
\noalign{\smallskip}\hline\noalign{\smallskip}
\multirow{2}{*}{Superficial}    & C             & (0,0,1)      & -1.0    & 0.05   & 0.022 & 0.01 & 1.0     & \multicolumn{1}{c}{-}  & 0.1  & -1.0 & 1.0   & \begin{tabular}[c]{@{}l@{}}0.5\\ 1.0\end{tabular} \\ \cline{2-13}
                                & T             & (1,0,0)      & -1.0    & 0.05   & 0.47  & 1.5  & 1.0     & \multicolumn{1}{c}{-}  & 1.0  & \multicolumn{1}{c}{-}  & \multicolumn{1}{c}{-} & \multicolumn{1}{c}{-} \\ \hline
\multirow{3}{*}{Middle}         & C             & (0,0,1)      & 0.0     & 0.05   & 0.022 & 2.0  & 1.0     & 1.0     & \multicolumn{1}{c}{-}  & 1.0 & 1.0   & 1.0      \\ \cline{2-13}
                                & T             & (0,0,1)      & 0.0     & 0.05   & 0.53  & 0.1  & 1.0     & 1.0     & \multicolumn{1}{c}{-}  & -1.0 & 1.0   & \begin{tabular}[c]{@{}l@{}}1.0\\ 2.0\end{tabular}  \\  \hline
\begin{tabular}[c]{@{}l@{}}Deep \end{tabular}           & T             & (0,0,1)      & 1.0     & 0.05   & 0.135  & 0.01 & 1.0     & 1.0     & \multicolumn{1}{c}{-}  & 1.0  & 1.0   & 1.0       \\ 
\noalign{\smallskip}\hline
\end{tabular}
\end{table*}
For each case, the structure parameters $k_1$ and $k_2$ were estimated. These parameters are not well understood for cartilage \citep{pierce2013hyperelastic} but they may change with the depth of the tissue. The same is shown with $\alpha$ and $\beta$ parameters. These new structure parameters introduced in this proposed model were defined as constant for all cases, except for compression in the superficial zone, where $\beta$ was defined as $0.1$ to adjust better the tension-compression curve to experimental data. A convergence problem in the numerical example led to use a different value of $\beta$. To better understand the influence of $\alpha$ and $\beta$ structure parameters, some representative numerical examples were performed (see Sect.~\ref{results1}). \\
Table \ref{tab:2} gives the set of material and structure parameters used in the proposed fibre-reinforced FE model to estimate the nonlinear response of the three cartilage layers (superficial, middle and deep zones) under compression (C) and tension (T) conditions.
\subsection{Remodelling analysis \label{remodellinganalysis}}
The remodelling algorithm was integrated in the fibre-reinforced FE model. Some numerical examples using the proposed approach were performed for all zones of cartilage (superficial, middle and deep). To illustrate the major features of the fibre remodelling model, the cubic sample was loaded axially (in tension and compression) and the axial true stress was analysed. For each zone, the material and structure values were assumed the same as in the previous examples (see Table \ref{tab:2}). Initially, the remodelling parameter $r_b$ was arbitrary set to $1.0$. In parallel with simulations, its influence in the fibre distribution parameter was analysed. As the value of the constant rate $\kappa$ showed to be not critical for the evolution of the remodelling process and as there is no experimental data to validate this parameter, it was set $1.0$ for all simulations. The remodelling process was considered with a dimensionless time scale. The load was applied instantaneously, thereafter it is held constant and the remodelling process starts. The evolution of the fibre orientation and stress fields are analysed during the remodelling process, as shown in Fig. 14-16. The remodelling parameters used in all simulations are displayed in Table \ref{tab:2}.

\section{Results and Discussion \label{results}}
\subsection{Tension-compression nonlinear behaviour \label{results1}}
\begin{figure}[htbp]
	\centering
		\includegraphics[width=0.50\textwidth]{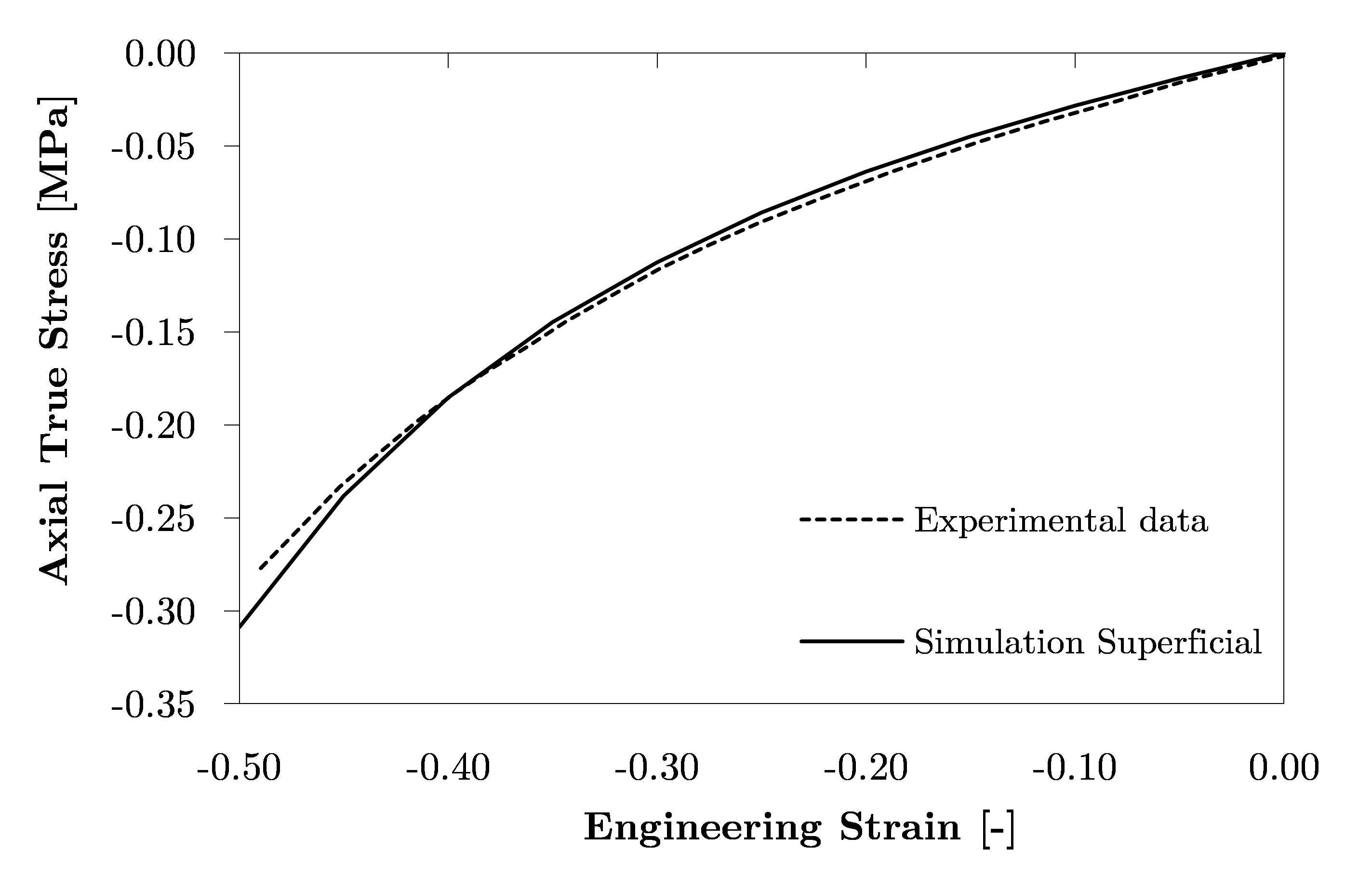}
	\caption{Unconfined compression stress of the FE simulation for the superficial zone and comparison with the experimental data from \citet{ateshian1997finite}.}
	\label{fig:fig5}
\end{figure}

After an extensive exploration for data which might validate our FE model, some literature data \citep{ateshian1997finite,elliott2002direct,jurvelin2003mechanical,pierce2013hyperelastic} was selected. 
A uniaxial unconfined compression test was carried out to demonstrate the validation of the constitutive model for the superficial zone (with the fibres aligned in the loading direction). Fig. \ref{fig:fig5} shows a comparison between the computed axial Cauchy stress component obtained by the FE model and compared with the experimental results available \citep{ateshian1997finite}. Simulation coincides well within one standard deviation of the stresses determined experimentally, reproducing the mechanical behaviour of the superficial zone (distribution defined as $b=-1$). When the cuboic FE model is loaded, the resulting axial stress increases (in absolute value) nonlinearly with the stretch. The FE model validation required a little effort. Although, a difference between a confined and unconfined compression was not considered in this study and it is believed that probably this different is associated with the hard validation. To achieve a better adjustment to the experimental results, and as the nonlinearity of the stress-strain curves can be controlled with the structural parameters, a different value of $\beta$ parameter ($0.1$ instead of $1.0$) was defined in this simulation. The analysis of $\beta$ parameter is presented below. 

\begin{figure}[htbp]
	\centering
		\includegraphics[width=0.50\textwidth]{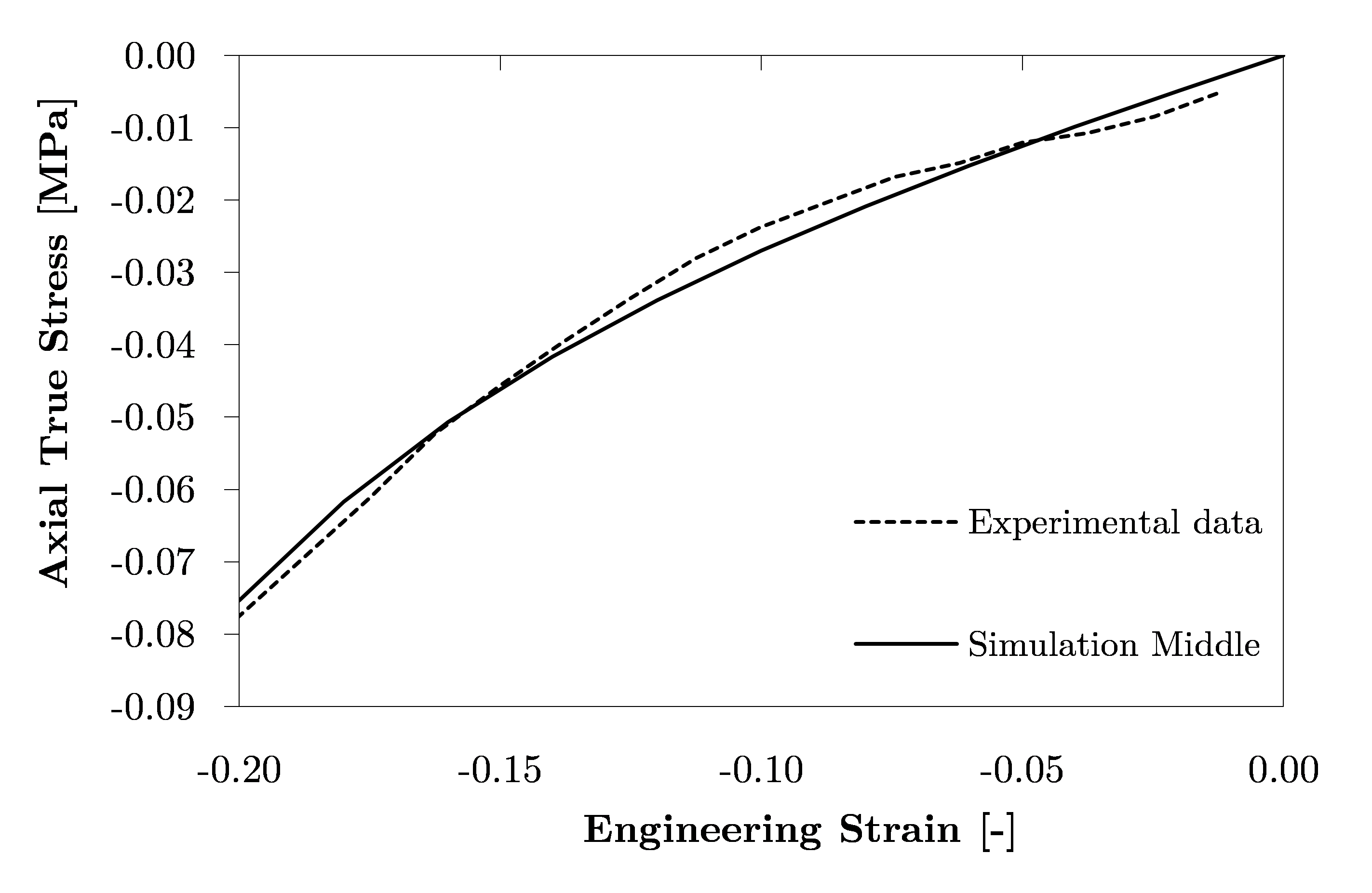}
	\caption{Unconfined compression stress of the FE simulation for the middle zone and comparison with the experimental data from \citet{jurvelin2003mechanical}.}
	\label{fig:fig6}
\end{figure}

An FE model of the middle cartilage zone with a random fibre orientation was study using the experimental data from \citet{jurvelin2003mechanical}. \\
Fig. \ref{fig:fig6} shows the computed axial true stress versus engineering strain for both simulation model and experimental data. The proposed fibre-reinforced FE model fits the nonlinear behaviour (in agreement with standard deviations from experimental results) and it is able to reproduce the experimental data in unconfined compression conditions.

\begin{figure}[htbp]
	\centering
		\includegraphics[width=0.50\textwidth]{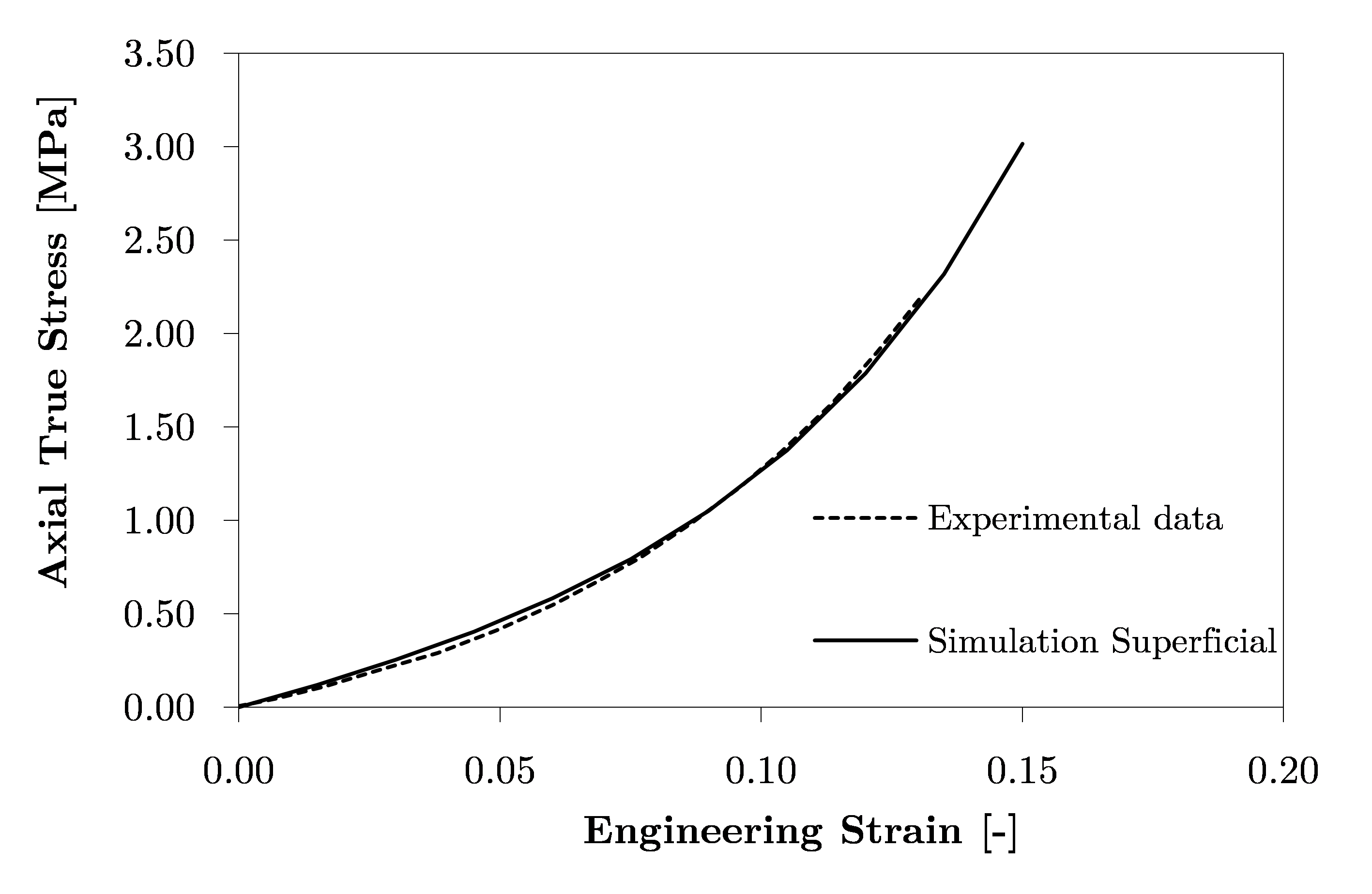}
	\caption{Uniaxial tension stress of the FE simulation for superficial zone of cartilage and comparison with the experimental data from \citet{elliott2002direct}.}
	\label{fig:fig7}
\end{figure}

\begin{figure}[htbp]
	\centering
		\includegraphics[width=0.50\textwidth]{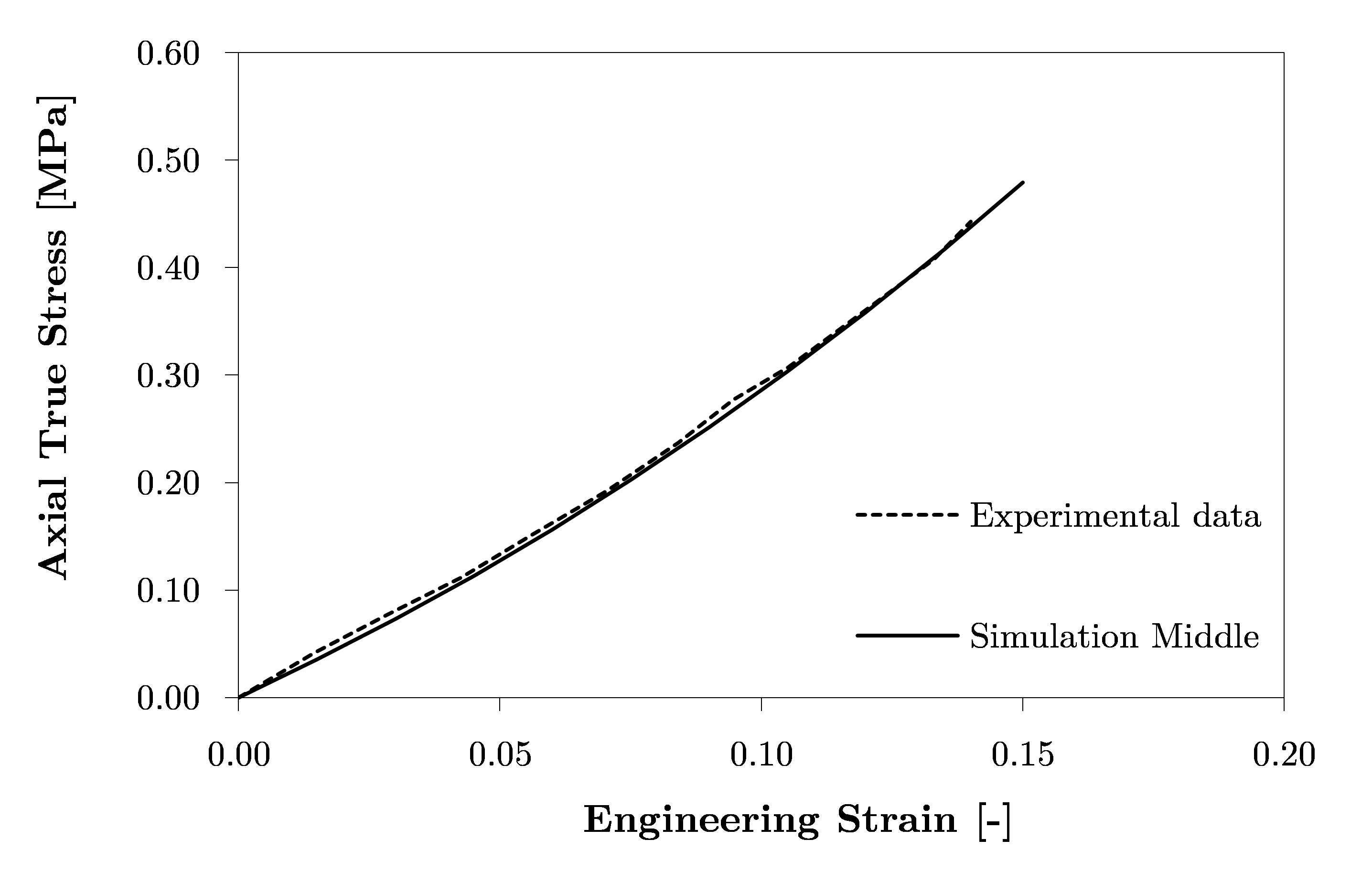}
	\caption{Uniaxial tension stress of the FE simulation for middle zone of cartilage and comparison with the experimental data from \citep{elliott2002direct}.}
	\label{fig:fig8}
\end{figure}

The uniaxial tension simulations of the superficial and middle zones are presented in Fig. \ref{fig:fig7} and Fig. \ref{fig:fig8}, respectively. In the first simulation, all fibres are defined in the plane parallel (perpendicular to the fibre reference direction) to the loading direction. The results exactly fit the experiments \citep{elliott2002direct} for tensile testing of this zone very accurately. Similarly, the tensile results for the middle zone overlap the experimental data from \citet{elliott2002direct}. 


\begin{figure}[htbp]
	\centering
		\includegraphics[width=0.50\textwidth]{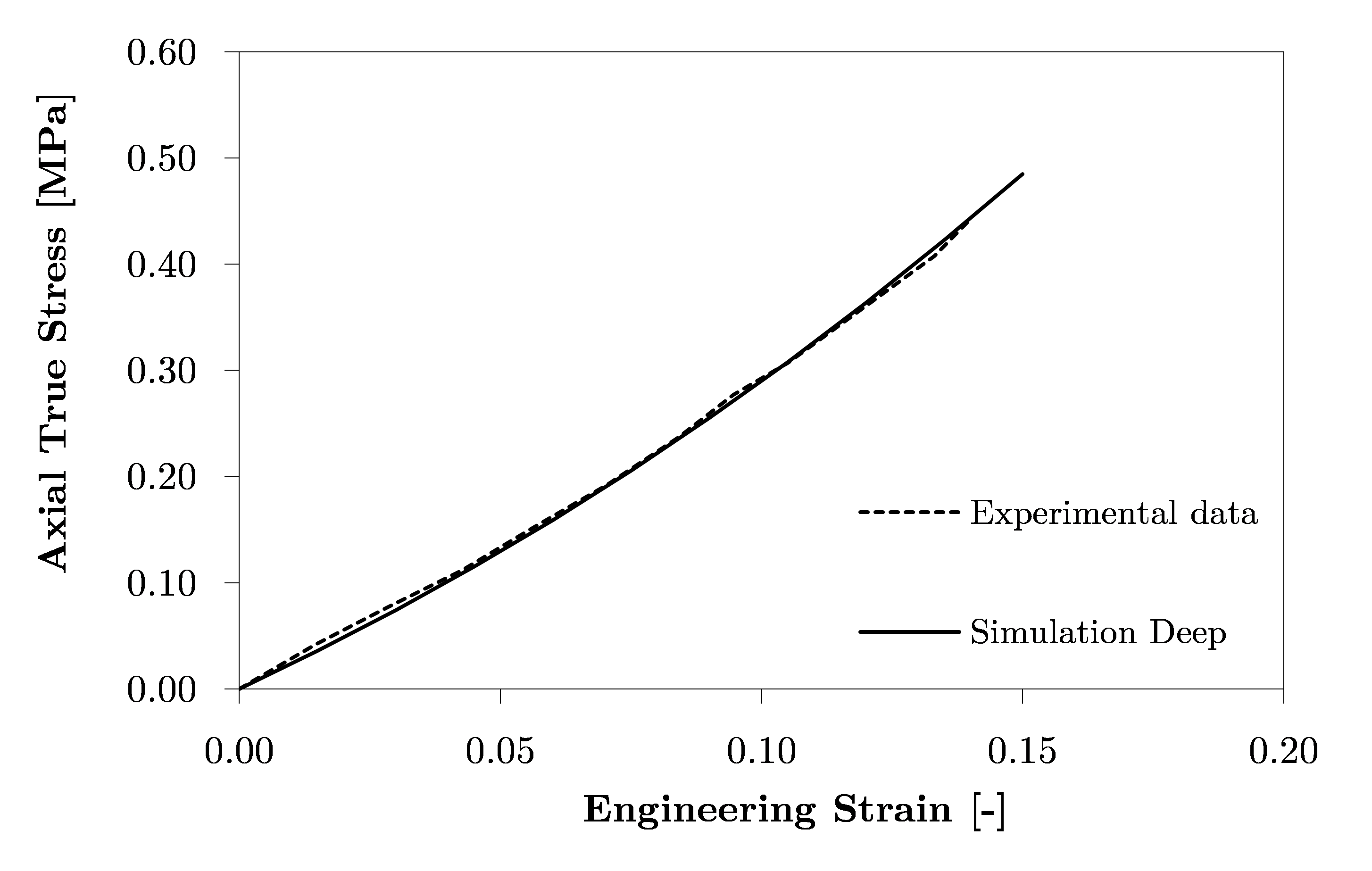}
	\caption{Uniaxial tension stress of the FE simulation for deep zone of cartilage and validation with data of the middle zone stress-strain curve from \citet{elliott2002direct}. In this case, the parameters should be adjusted.}
	\label{fig:fig9}
\end{figure}

To demonstrate that the proposed fibre reinforced FE model can be used to simulate the cartilage deep zone, a representative unconfined compression test (using the experimental data from the middle zone \citep{elliott2002direct} was performed. Fig. \ref{fig:fig9} shows the results of this representative simulation. Although the experimental data in this case are not available, the model proved to be suitable in cases where the fibres are perfectly aligned ($b=1$). For feasible experimental data of the deep zone, the model parameters should be adjusted.
\begin{figure}[htbp]
	\centering
		\includegraphics[width=0.50\textwidth]{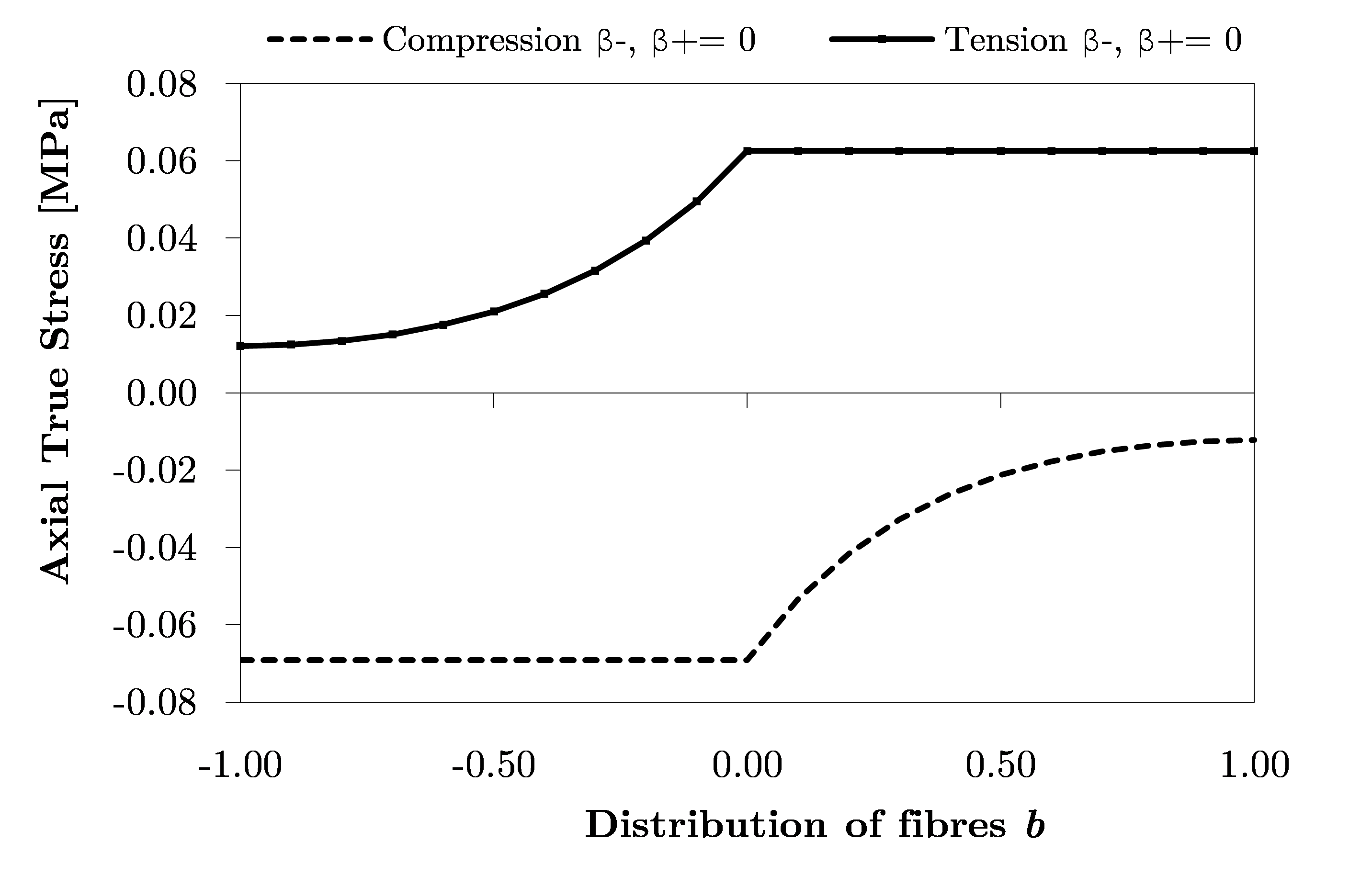}
	\caption{The influence of structure parameter on stress response for different fibres distributions under compression and tension simulations.}
	\label{fig:fig10}
\end{figure}

\paragraph{The role of structure parameters $\alpha$ and $\beta$}
\ The influence of $\alpha$ and $\beta$ parameters in the tension-compression response was also investigated. 
These parameters are used to complement the structure tensor increasing the flexibility of the model. A representative numerical example was performed and the relationship between the fibre distribution ($b$) and the variation of the stress with the $\beta$ parameter is plotted in Fig. \ref{fig:fig10}. In all cases, the parameter $\alpha$ was kept constant as $1.0$. For $\beta$ ($+$ and $-$) defined as zero, the stress increases in tension and it is constant in compression for negative values of $b$ (superficial zone) until reach the totally isotropic region (middle zone with $b=0$). Consequently, for a positive value of $b$ the behaviour is opposite, the axial stress is constant in tension and increases in compression (from the isotropic distribution to the distribution of perfectly fibres aligned).

\begin{figure}[htbp]
	\centering
		\includegraphics[width=0.50\textwidth]{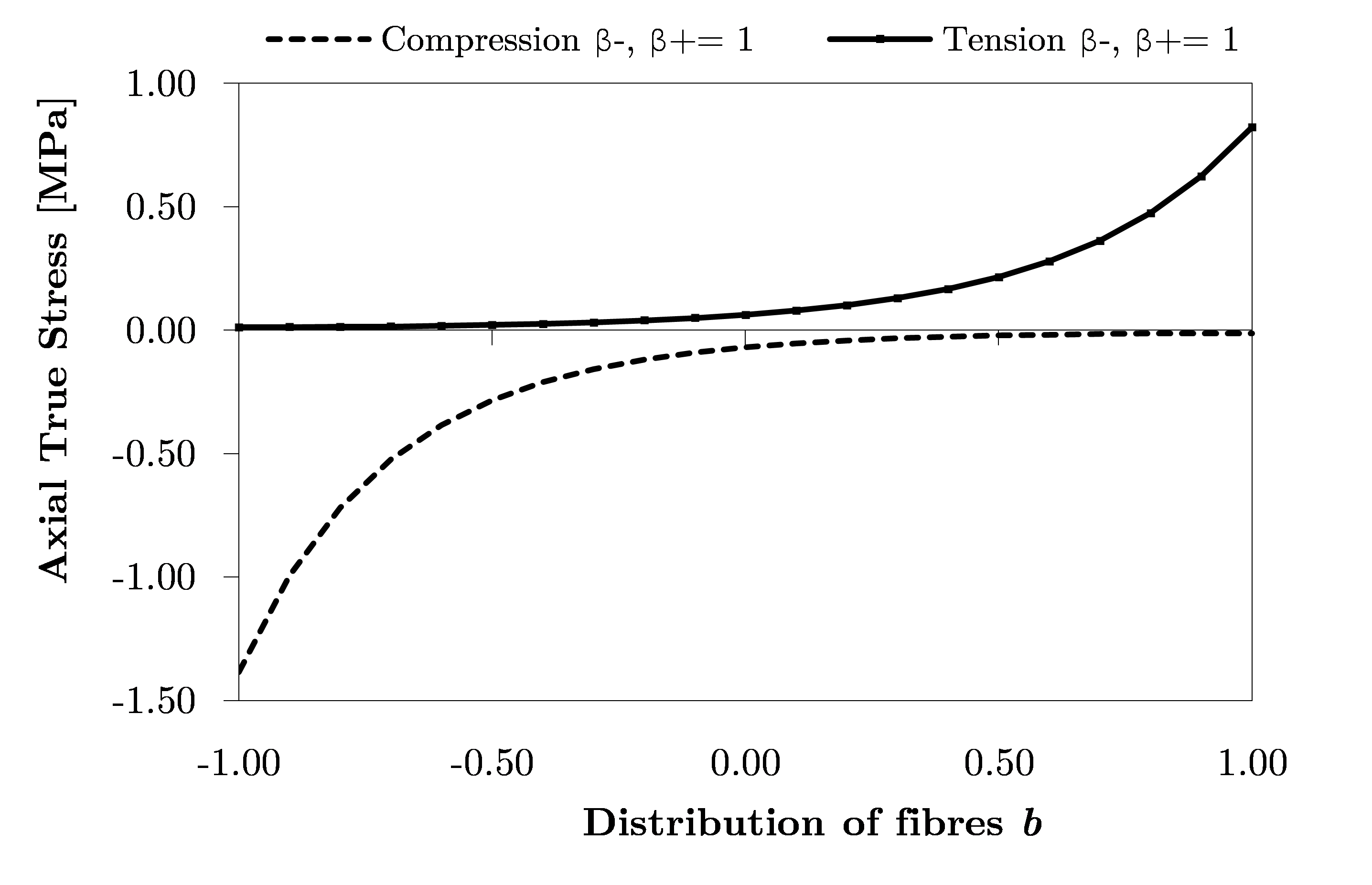}
	\caption{The influence of structure parameter $\beta=1$ on stress response for different fibres distributions under compression and tension simulations.}
	\label{fig:fig11}
\end{figure}

\begin{figure}[htbp]
	\centering
		\includegraphics[width=0.50\textwidth]{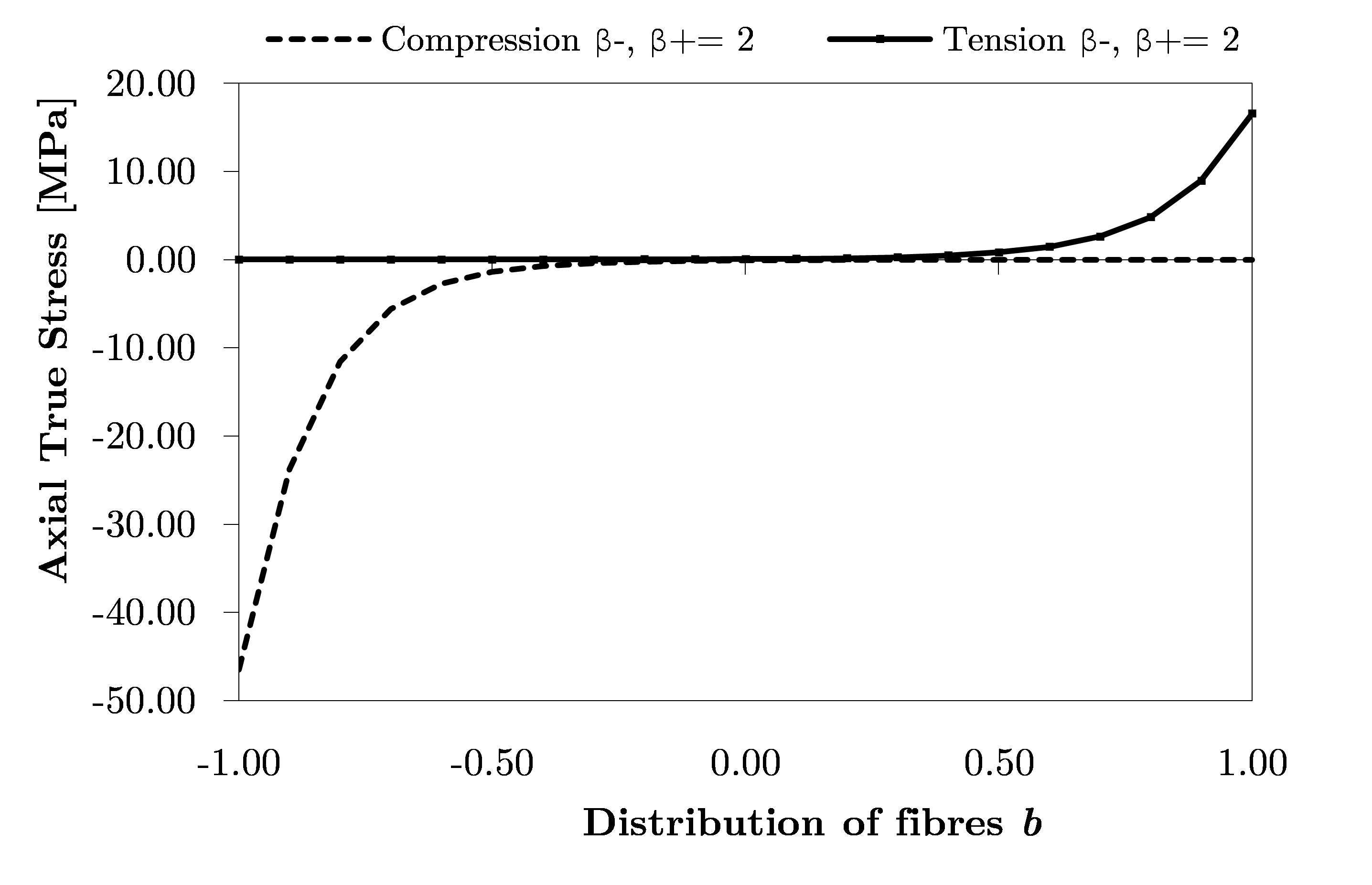}
	\caption{The influence of structure parameter $\beta=2$ on stress response for different fibres distributions under compression and tension simulations.}
	\label{fig:fig12}
\end{figure}

When $\beta$ ($+$ and $-$) is considered in the structure tensor for values different from zero, stress increases in tension from the superficial zone (where $b$ is more negative) to the deep zone ($b=-1$) (see Fig. \ref{fig:fig11}). The opposite occurs when the model is under compression. If the $\beta$ parameter increases (see Fig. \ref{fig:fig12}), the maximum value of the stress will increase in compression and also in tension.

\begin{figure}[htbp]
	\centering
		\includegraphics[width=0.50\textwidth]{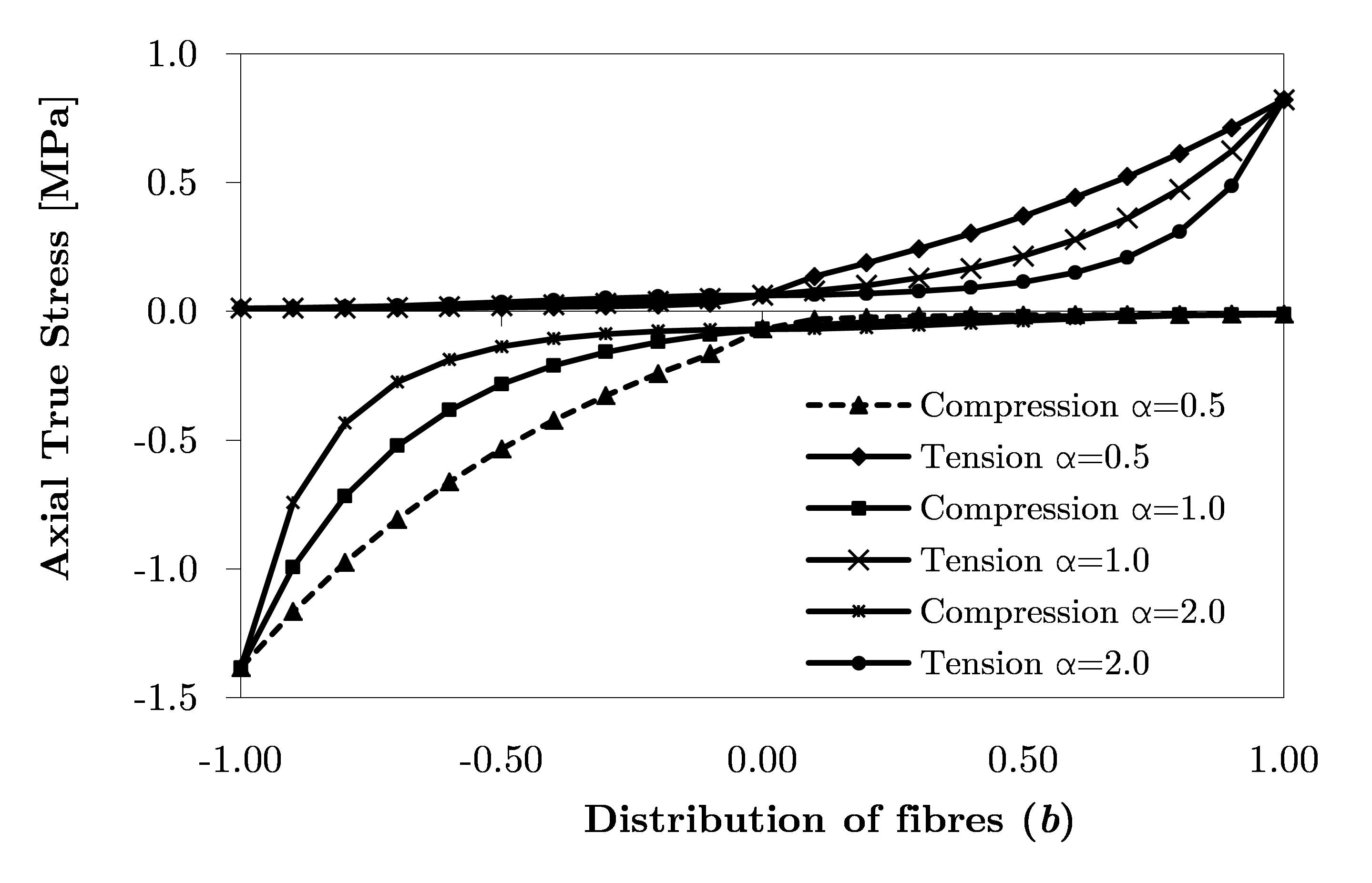}
	\caption{The influence of structure parameter $\alpha$ on the stress response for different fibres distributions under compression and tension simulations.}
	\label{fig:fig13}
\end{figure}

Another parameter which characterizes the structure tensor is the parameter $\alpha$, which allows the adjustment of the stress response curvature as function of the fibre alignment. As shown in Fig. \ref{fig:fig13}, the change of this parameter gives a different result in the stress response.
\subsection{Remodelling process \label{remodellingprocess}}
\begin{figure}[htbp]
	\centering
		\includegraphics[width=0.50\textwidth]{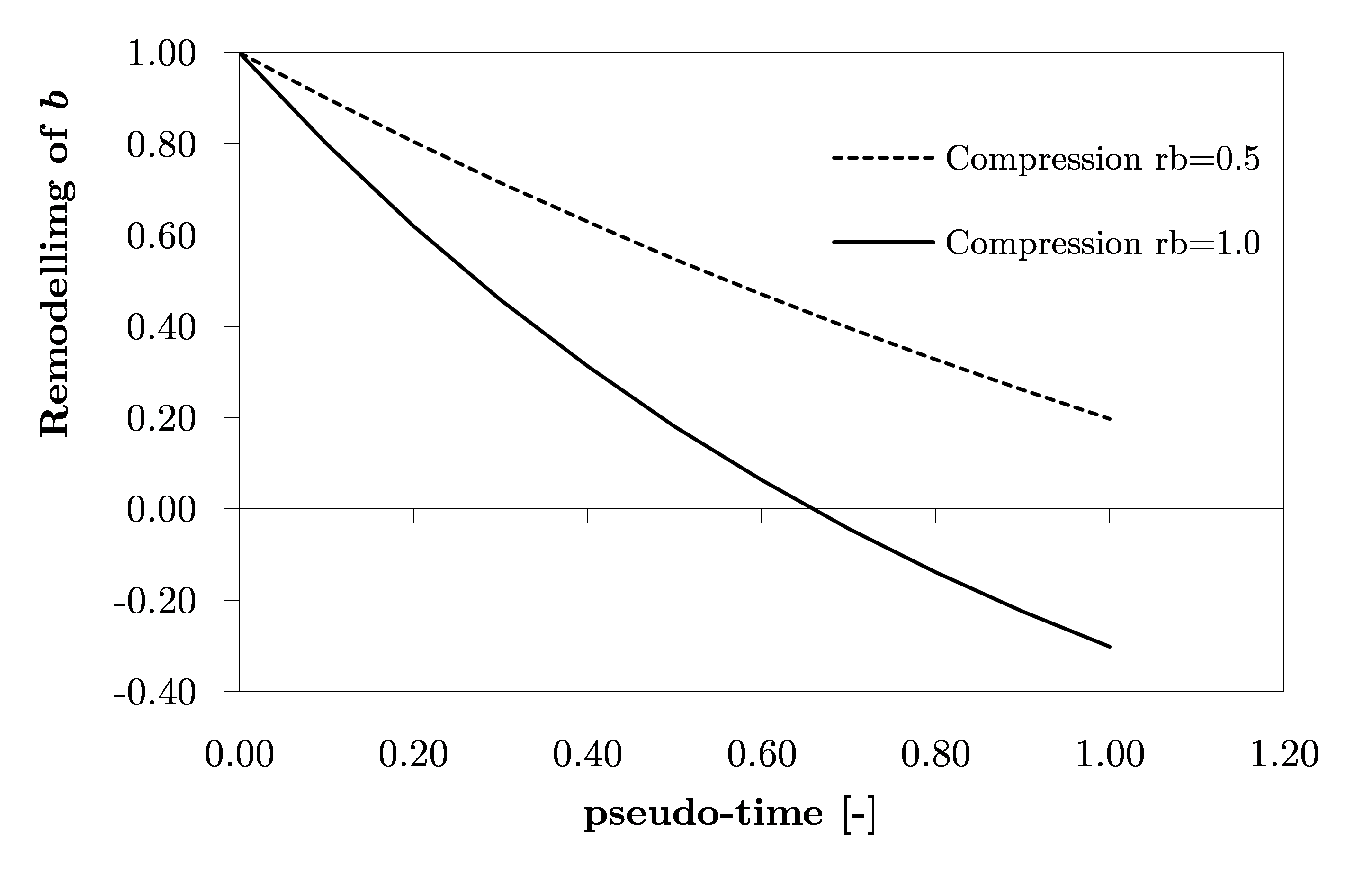}
	\caption{Remodelling of fibre distribution versus time in compression with a reference fibre direction ${\vec e_{f,0}} = (1,0,0)$ and $b_0=1$ (representing an unknown zone). The dotted line with a $r_b=0.5$ and the continuous line with $r_b=1.0$.}
	\label{fig:fig14}
\end{figure}

\begin{figure}[htbp]
	\centering
		\includegraphics[width=0.50\textwidth]{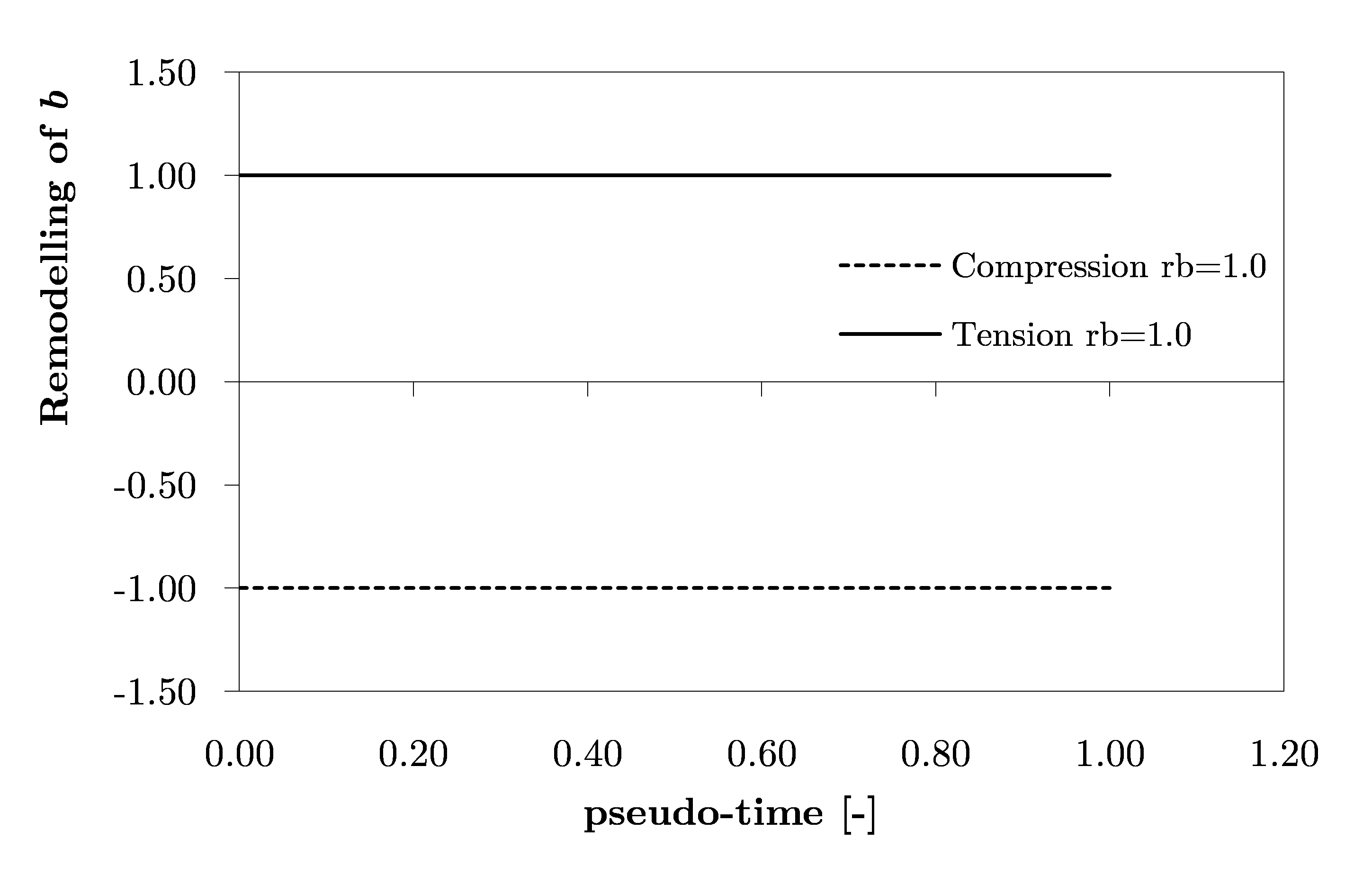}
	\caption{Remodelling of fibre distribution versus time in compression with a reference fibre direction ${\vec e_{f,0}} = (0,0,1)$ and ${b_0}=-1$ (representing the superficial zone).Remodelling of fibre distribution versus time in tension with a reference fibre direction ${\vec e_{f,0}} = (0,0,1)$ and ${b_0}=1$ (representing the deep zone).}
	\label{fig:fig15}
\end{figure}

\begin{figure}[htbp]
	\centering
		\includegraphics[width=0.50\textwidth]{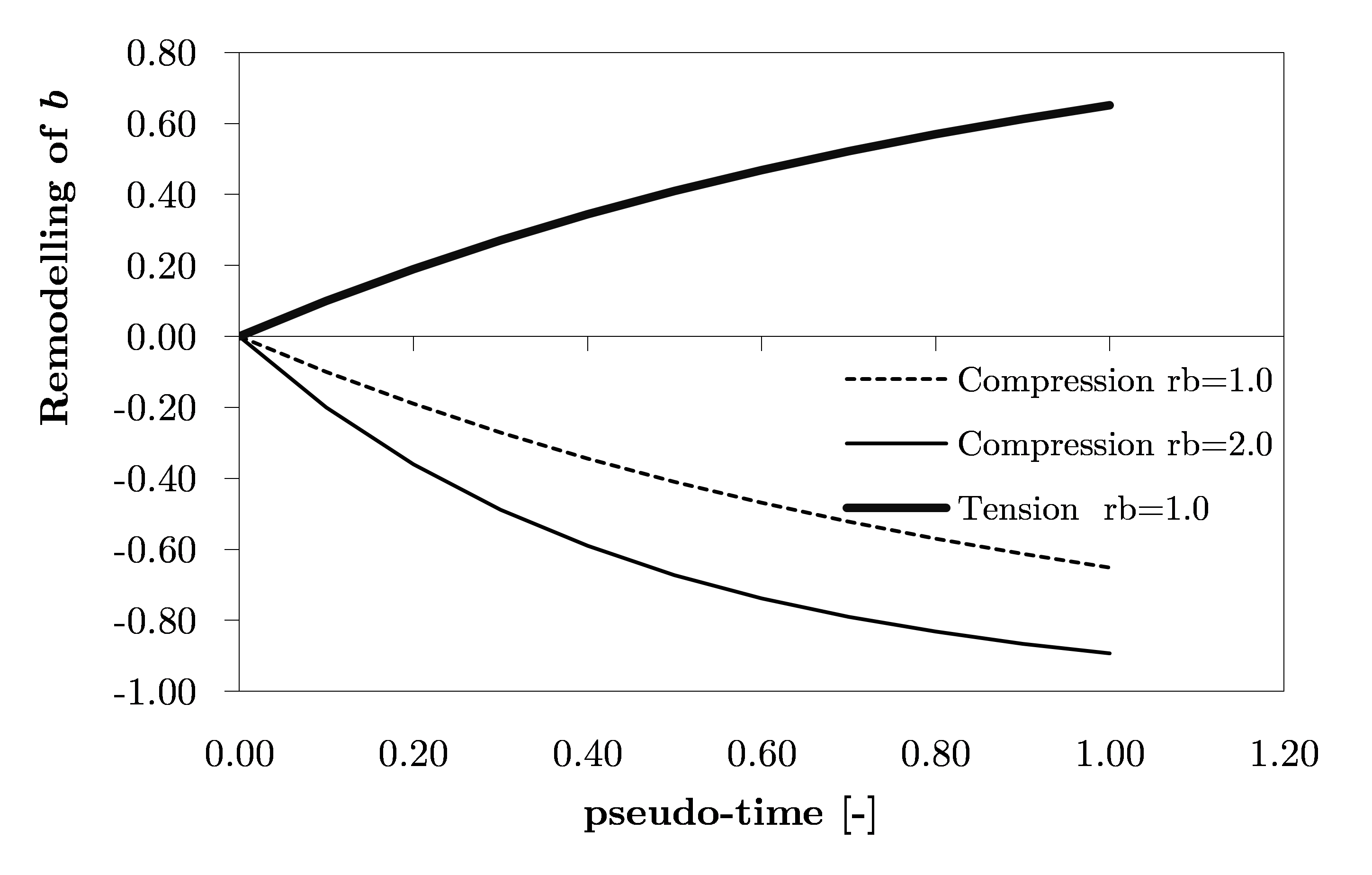}
	\caption{Remodelling of fibre distribution versus time in compression and tension with a reference fibre direction ${\vec e_{f,0}} = (0,0,1)$ and ${b_0}=0$ (representing the middle zone).}
	\label{fig:fig16}
\end{figure}

\begin{figure*}[htbp]
	\centering
		\includegraphics[width=1.0\textwidth]{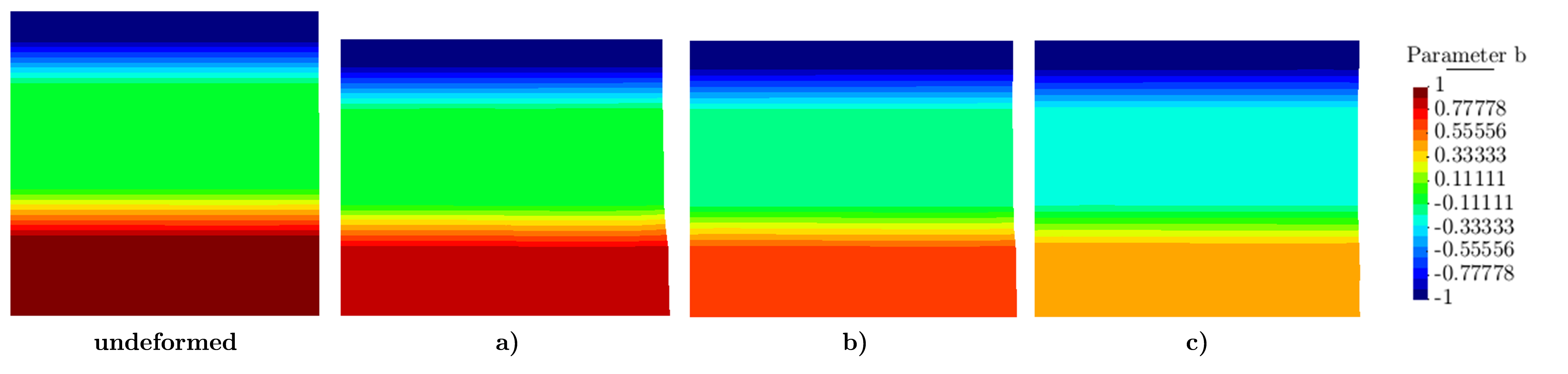}
	\caption{Compression simulation of a three-layer FE model with different fibres distribution and three different remodelling parameters a)$r_b=0.1$ b)$r_b=0.2$ and c)$r_b=0.3$. }
	\label{fig:fig17}
\end{figure*}

To evaluate the new remodelling approach, the evolution of the fibre distribution parameter and the fibre reorientation under different loading conditions was investigated. Fig. \ref{fig:fig14}$-$\ref{fig:fig17} compares the evolution of $b$  in unconfined compression versus pseudo-time stepping, with a different value for $r_b$ parameter. Both simulations show (Fig. \ref{fig:fig14}) that the distribution of fibres tends to negative values, as it should, since that all fibres, initially aligned (${b_0}=1$) with the reference direction ${\vec e_{f,0}} = (1,0,0)$, tend to be distributed isotopically in the plane perpendicular to the loading direction, and the reference fibre vector becomes perpendicular to this reference direction. However, an increase of $r_b$ from $0.5$ to $1.0$ leads to the $b$ parameter reaches faster the expected value ${b_t}=-1$. It should be noted that this case does not correspond to any point of the cartilage and it is only referred to a better understand the presented model.
Fig. \ref{fig:fig15} shows the remodelling of $b$ for a reference fibre direction of ${\vec e_{f,0}} = (0,0,1)$, when fibres are organized following the plane perpendicular to this direction (${b_0}=-1$). Thus, it represents the superficial zone of the cartilage. When this zone is compressed, the distribution of fibres (${b_t}=-1$) tends to be constant along the simulation. There is no remodelling process of fibres because in compression they will expand in the same plane and no change in the directions is observed. The same occurs when the distribution of fibres is aligned with the reference direction. The distribution was kept as constant ($b=1$) during the simulation (see Fig. \ref{fig:fig15}). This result can show the natural phenomenon that occurs with the fibres in the deep zone. When the tissue is loaded in tension, all fibres aligned with the reference direction tend to stretch (along the axial direction) and keeping the same direction while they are in tension. In compression, fibres in this layer of the tissue do not respond, resulting in a phenomenon called ‘buckling’ of fibres \citep{federico2008towards,cortes2010characterizing}. Thereby, only the tension simulation was presented in this study.
To analyse the evolution of the isotropic distribution, which is associated with the middle zone (${b_0}=0$) of the tissue, one model in tension and two models in unconfined compression were investigated. The fibre reference direction was defined by ${\vec e_{f,0}} = (0,0,1)$ in both simulations. The results of these three simulations are presented in Fig. \ref{fig:fig16}. 
In tension, ${b_0}$ is initialized as zero and tends to positive values (where the limit is ${b_t}=1$). This reorientation of fibres indicates that, in tension, all randomly oriented fibres tend to realignment with the loading direction, which is the same of the reference fibre direction. In compression, the distribution of fibres varied between ${b_0}=0$ and ${b_t}=-1$. Initially, fibres show an isotropic distribution (middle zone) and then, they tend to be perpendicular to the reference direction ${\vec e_{f,0}} = (0,0,1)$ but with an isotropic distribution in the plane (${b_t}=-1$). As shown in a previous numerical example (Fig. \ref{fig:fig14}), when the $r_b$ increases the negative limit (${b_t}=-1$) is reached faster (${r_b}=2$). These results highlighted the assumptions initially made in this work (see Sect.~\ref{kinematics}). In compression the distribution of fibres parameter tends to be $-1$, characterizing a plane isotropic distribution and in tension the distribution parameter tends to be $1$, describing the perfect alignment of fibres with the reference direction \citep{driessen2003remodelling,wilson2006prediction}.
\paragraph{FE model with three layers}
\ To explore the influence of the distribution of fibres parameter (${b}$) under deformation with a depth-dependent manner, a combination of three layers in the same FE model was performed. The FE mesh used previously was divided into three different materials with 20$\%$, 50$\%$ and 30$\%$ of the total thickness for superficial, middle and deep zone, respectively. The fibre distribution for each zone was defined as ${b}=-1.0$ (superficial), ${b}=0.0$ (middle) and ${b}=1.0$ (deep) with a fibre reference direction ${\vec e_{f,0}} = (0,0,1)$. However, these ${b}$ values will change as function of tissue deformation history. The FE model was compressed by 10$\%$ of the sample thickness over a pseudo-time. The axial displacement was applied perpendicular to the top of superficial zone.  The results of these simulations are presented in Fig.\ref{fig:fig17} with different values for the remodelling parameter a)$r_b=0.1$ b)$r_b=0.2$ and c)$r_b=0.3$. 
For the same time step, the remodelling parameter ($r_b$) influences the distribution of fibres (${b}$). With deformation, fibres run to the direction which can resist more. As previously discussed, when the sample is compressed, fibres reorient to have a distribution of ${b}=-1.0$. This phenomenon is evident in the middle and deep zones, while in the superficial zone it keeps constant along the time. The remodelling parameter controls the rate of change of the distribution of fibres. When the $r_b$ value increases, the remodelling process is faster to the new direction, determined from the deformation history. It is shown Fig.\ref{fig:fig17} where the distribution in the deep zone is ${b}\approx0.8$ for $r_b=0.1$ (Fig.\ref{fig:fig17}a), ${b}\approx0.6$ for $r_b=0.2$ (Fig.\ref{fig:fig17}b) and ${b}\approx0.3$ for $r_b=0.3$ (Fig.\ref{fig:fig17}c). 
\section{Conclusion \label{conclusion}}
In the present study, a novel continuum anisotropic hyperelastic remodelling approach to investigate the effects of the cartilage collagen network in a depth-dependent manner was proposed. A new structure tensor, in order to describe the mechanical response of the embedded anisotropic collagen fibres, was introduced and different material and structure parameters were included to increase the flexibility of the new proposed model. The preliminary numerical results allow to demonstrate the ability of the new formulation to reproduce the nonlinear tension-compression response of cartilage in two main loading conditions, and cross validated with the experimental data available from the literature. In the superficial zone, the validation of these parameters was more complex either by the absence of experimental data or by the confined configuration used in the current experiments. In case of the shear modulus of the isotropic matrix, the value was assumed as equal in all regions being low when compared with literature.  \\
The proposed FE model showed good results and low numerical consumption when compared with other approaches in literature. This formulation is relatively simple when compared with the numerical integration on the unit sphere \citep{federico2008towards} where the spherical designs methods are used. In the present model the dispersion of fibres was considered in the hyperelastic formulation through a structure tensor, which gives a different mechanical response when compared with models where the anisotropic strain depends only by the local reference direction of the collagen fibres \citep{holzapfel2010constitutive,pierce2013hyperelastic,pierce2015microstructurally}. Moreover, the buckling phenomenon of the fibres was also taken into account.\\
In the remodelling cases, the evolution of the collagen fibres orientation for each zone was analysed. Again, due to the lack on experimental data, the study of regeneration fibre parameter was not totally achieved. Thus, some remodelling parameters were fixed in all simulations. Although this work has focused on changes in fibres orientation, the effects of fibre remodelling in the mechanical properties of the tissue should be further investigated. Diverse fibre distributions could be explored using this model and more real-like boundary conditions used instead of the layer-based sample. \\
This new constitutive law for modelling the anisotropy, reorientation and remodelling behaviour of the articular cartilage can contribute to the development of the future studies in the articular cartilage. In addition to cartilage tissue, the present model has the potential to study other fibrous soft tissues, contributing to a better knowledge of the structural adaptation in the collagen network associated with arbitrary mechanical stimuli. In summary, this study aims to improve numerical models that have proven to be essential tools to support the experimental protocols in the cartilage tissue engineering.
\section{Conflict of interest \label{conflict}}
The authors declare that they have no conflict of interest.
\begin{acknowledgements}
The first author is grateful to FCT – $\textit{Fundação para a Ciência e a Tecnologia}$ (Portugal) for the PhD grant (SFRH/BD/87933/2012). This work is funded by FEDER through COMPETE - $\textit{Programa}$ $\textit{Operacional}$ $\textit{de}$ $\textit{Fatores}$ $\textit{de}$ $\textit{Competitividade}$ (COMPETE 2020-PTDC/EMS-TEC/3263/2014) and by national funds through FCT, under the strategic project PEst-C/EME/UI0481/2013 and also in the scope of the following project: FCOMP-01-0124-FEDER-015191.
\end{acknowledgements}

\bibliographystyle{spbasic}      

\bibliography{References}

\begin{thebibliography}{29}
\providecommand{\natexlab}[1]{#1}
\providecommand{\url}[1]{{#1}}
\providecommand{\urlprefix}{URL }
\expandafter\ifx\csname urlstyle\endcsname\relax
  \providecommand{\doi}[1]{DOI~\discretionary{}{}{}#1}\else
  \providecommand{\doi}{DOI~\discretionary{}{}{}\begingroup
  \urlstyle{rm}\Url}\fi
\providecommand{\eprint}[2][]{\url{#2}}

\bibitem[{Alves et~al(2010)Alves, Yamamura, Oda, Teodosiu, Middleton, Evans,
  Holt, Jacobs, Atienza, and Walker}]{alves2010numerical}
Alves J, Yamamura N, Oda T, Teodosiu C, Middleton J, Evans S, Holt C, Jacobs C,
  Atienza C, Walker B (2010) Numerical simulation of musculo-skeletal systems
  by v-biomech. Proceedings of 9th International Symposium CMBBE2010

\bibitem[{Ateshian et~al(1997)Ateshian, Warden, Kim, Grelsamer, and
  Mow}]{ateshian1997finite}
Ateshian G, Warden W, Kim J, Grelsamer R, Mow V (1997) Finite deformation
  biphasic material properties of bovine articular cartilage from confined
  compression experiments. Journal of biomechanics 30(11):1157--1164

\bibitem[{Ateshian et~al(2009)Ateshian, Rajan, Chahine, Canal, and
  Hung}]{ateshian2009modeling}
Ateshian GA, Rajan V, Chahine NO, Canal CE, Hung CT (2009) Modeling the matrix
  of articular cartilage using a continuous fiber angular distribution predicts
  many observed phenomena. Journal of biomechanical engineering 131(6):061,003

\bibitem[{Bandeiras et~al(2015)Bandeiras, Completo, and
  Ramos}]{bandeiras2015cartilage}
Bandeiras C, Completo A, Ramos A (2015) Influence of the scaffold geometry on
  the spatial and temporal evolution of the mechanical properties of
  tissue-engineered cartilage: insights from a mathematical model. Biomechanics
  and modeling in mechanobiology pp 1--14

\bibitem[{Castro et~al(2014)Castro, Wilson, Huyghe, Ito, and
  Alves}]{castro2014disc}
Castro A, Wilson W, Huyghe J, Ito K, Alves J (2014) Intervertebral disc creep
  behavior assessment through an open source finite element solver. Journal of
  biomechanics 47(1):297--301

\bibitem[{Chung and Ho(2010)}]{chung2010analysis}
Chung C, Ho SY (2010) Analysis of collagen and glucose modulated cell growth
  within tissue engineered scaffolds. Annals of biomedical engineering
  38(4):1655--1663

\bibitem[{Cortes et~al(2010)Cortes, Lake, Kadlowec, Soslowsky, and
  Elliott}]{cortes2010characterizing}
Cortes DH, Lake SP, Kadlowec JA, Soslowsky LJ, Elliott DM (2010) Characterizing
  the mechanical contribution of fiber angular distribution in connective
  tissue: comparison of two modeling approaches. Biomechanics and modeling in
  mechanobiology 9(5):651--658

\bibitem[{Driessen et~al(2003)Driessen, Peters, Huyghe, Bouten, and
  Baaijens}]{driessen2003remodelling}
Driessen N, Peters G, Huyghe J, Bouten C, Baaijens F (2003) Remodelling of
  continuously distributed collagen fibres in soft connective tissues. Journal
  of biomechanics 36(8):1151--1158

\bibitem[{Driessen et~al(2008)Driessen, Cox, Bouten, and
  Baaijens}]{driessen2008remodelling}
Driessen NJ, Cox MA, Bouten CV, Baaijens FP (2008) Remodelling of the angular
  collagen fiber distribution in cardiovascular tissues. Biomechanics and
  modeling in mechanobiology 7(2):93--103

\bibitem[{Elliott et~al(2002)Elliott, Narmoneva, and
  Setton}]{elliott2002direct}
Elliott DM, Narmoneva DA, Setton LA (2002) Direct measurement of the poisson
  ratio of human patella cartilage in tension. Journal of biomechanical
  engineering 124(2):223--228

\bibitem[{Federico and Gasser(2010)}]{federico2010nonlinear}
Federico S, Gasser TC (2010) Nonlinear elasticity of biological tissues with
  statistical fibre orientation. Journal of the Royal Society Interface
  7(47):955--966

\bibitem[{Federico and Herzog(2008)}]{federico2008towards}
Federico S, Herzog W (2008) Towards an analytical model of soft biological
  tissues. Journal of biomechanics 41(16):3309--3313

\bibitem[{Gasser et~al(2006)Gasser, Ogden, and
  Holzapfel}]{gasser2006hyperelastic}
Gasser TC, Ogden RW, Holzapfel GA (2006) Hyperelastic modelling of arterial
  layers with distributed collagen fibre orientations. Journal of the royal
  society interface 3(6):15--35

\bibitem[{Guo et~al(2015)Guo, Maher, and Torzilli}]{guo2015biphasic}
Guo H, Maher SA, Torzilli PA (2015) A biphasic finite element study on the role
  of the articular cartilage superficial zone in confined compression. Journal
  of biomechanics 48(1):166--170

\bibitem[{Holzapfel and Gasser(2001)}]{holzapfel2001viscoelastic}
Holzapfel GA, Gasser TC (2001) A viscoelastic model for fiber-reinforced
  composites at finite strains: Continuum basis, computational aspects and
  applications. Computer methods in applied mechanics and engineering
  190(34):4379--4403

\bibitem[{Holzapfel and Ogden(2010)}]{holzapfel2010constitutive}
Holzapfel GA, Ogden RW (2010) Constitutive modelling of arteries. Proceedings
  of the Royal Society of London A: Mathematical, Physical and Engineering
  Sciences 466(2118):1551--1597

\bibitem[{Holzapfel et~al(2000)Holzapfel, Gasser, and Ogden}]{holzapfel2000new}
Holzapfel GA, Gasser TC, Ogden RW (2000) A new constitutive framework for
  arterial wall mechanics and a comparative study of material models. Journal
  of elasticity and the physical science of solids 61(1-3):1--48

\bibitem[{Hossain et~al(2014)Hossain, Bergstrom, and
  Chen}]{hossain2014prediction}
Hossain MS, Bergstrom D, Chen X (2014) Prediction of cell growth rate over
  scaffold strands inside a perfusion bioreactor. Biomechanics and modeling in
  mechanobiology 14(2):333--344

\bibitem[{Jurvelin et~al(2003)Jurvelin, Buschmann, and
  Hunziker}]{jurvelin2003mechanical}
Jurvelin J, Buschmann M, Hunziker E (2003) Mechanical anisotropy of the human
  knee articular cartilage in compression. Proceedings of the Institution of
  Mechanical Engineers, Part H: Journal of Engineering in Medicine
  217(3):215--219

\bibitem[{Khoshgoftar et~al(2011)Khoshgoftar, van Donkelaar, and
  Ito}]{khoshgoftar2011mechanical}
Khoshgoftar M, van Donkelaar CC, Ito K (2011) Mechanical stimulation to
  stimulate formation of a physiological collagen architecture in
  tissue-engineered cartilage: a numerical study. Computer methods in
  biomechanics and biomedical engineering 14(02):135--144

\bibitem[{Khoshgoftar et~al(2013)Khoshgoftar, Wilson, Ito, and van
  Donkelaar}]{khoshgoftar2013effect}
Khoshgoftar M, Wilson W, Ito K, van Donkelaar CC (2013) The effect of
  tissue-engineered cartilage biomechanical and biochemical properties on its
  post-implantation mechanical behavior. Biomechanics and modeling in
  mechanobiology 12(1):43--54

\bibitem[{Nava et~al(2013)Nava, Raimondi, and
  Pietrabissa}]{nava2013multiphysics}
Nava MM, Raimondi MT, Pietrabissa R (2013) A multiphysics 3d model of tissue
  growth under interstitial perfusion in a tissue-engineering bioreactor.
  Biomechanics and modeling in mechanobiology 12(6):1169--1179

\bibitem[{Pearle et~al(2005)Pearle, Warren, and Rodeo}]{pearle2005basic}
Pearle AD, Warren RF, Rodeo SA (2005) Basic science of articular cartilage and
  osteoarthritis. Clinics in sports medicine 24(1):1--12

\bibitem[{Pierce et~al(2013)Pierce, Ricken, and
  Holzapfel}]{pierce2013hyperelastic}
Pierce DM, Ricken T, Holzapfel GA (2013) A hyperelastic biphasic
  fibre-reinforced model of articular cartilage considering distributed
  collagen fibre orientations: continuum basis, computational aspects and
  applications. Computer methods in biomechanics and biomedical engineering
  16(12):1344--1361

\bibitem[{Pierce et~al(2015)Pierce, Unterberger, Trobin, Ricken, and
  Holzapfel}]{pierce2015microstructurally}
Pierce DM, Unterberger MJ, Trobin W, Ricken T, Holzapfel GA (2015) A
  microstructurally based continuum model of cartilage viscoelasticity and
  permeability incorporating measured statistical fiber orientations.
  Biomechanics and modeling in mechanobiology pp 1--16

\bibitem[{Responte et~al(2007)Responte, Natoli, and
  Athanasiou}]{responte2007collagens}
Responte DJ, Natoli RM, Athanasiou KA (2007) Collagens of articular cartilage:
  structure, function, and importance in tissue engineering. Critical Reviews
  in Biomedical Engineering 35(5)

\bibitem[{Wilson et~al(2004)Wilson, Van~Donkelaar, Van~Rietbergen, Ito, and
  Huiskes}]{wilson2004stresses}
Wilson W, Van~Donkelaar C, Van~Rietbergen B, Ito K, Huiskes R (2004) Stresses
  in the local collagen network of articular cartilage: a poroviscoelastic
  fibril-reinforced finite element study. Journal of biomechanics
  37(3):357--366

\bibitem[{Wilson et~al(2006)Wilson, Driessen, Van~Donkelaar, and
  Ito}]{wilson2006prediction}
Wilson W, Driessen N, Van~Donkelaar C, Ito K (2006) Prediction of collagen
  orientation in articular cartilage by a collagen remodeling algorithm.
  Osteoarthritis and Cartilage 14(11):1196--1202

\bibitem[{Wilson et~al(2007)Wilson, Huyghe, and
  Van~Donkelaar}]{wilson2007depth}
Wilson W, Huyghe J, Van~Donkelaar C (2007) Depth-dependent compressive
  equilibrium properties of articular cartilage explained by its composition.
  Biomechanics and modeling in mechanobiology 6(1-2):43--53

\end{thebibliography}

%
%

\end{document}